\documentclass[useAMS,usenatbib]{mn2e}
\usepackage{myaasmacros,graphicx}

\def\ltsima{$\; \buildrel < \over \sim \;$}
\def\simlt{\lower.5ex\hbox{\ltsima}}   
\def\gtsima{$\; \buildrel > \over \sim \;$}
\def\simgt{\lower.5ex\hbox{\gtsima}}
\newcommand\bcite[1]{\citeauthor{#1} \citeyear{#1}}
\newcommand{\be}{\begin{equation}}
\newcommand{\ee}{\end{equation}}

\title[On the formation of dwarf galaxies and stellar halos]
{On the formation of dwarf galaxies and stellar halos}

\author[Read et. al.]{J. I. Read $^1$\thanks{Email: justin@physik.unizh.ch}, A. P. Pontzen$^2$ \& M. Viel$^{2,3}$
\\$^1$Institute of Theoretical Physics, University of Z\"urich,  
Wintherurestrasse 190, 8057 Zurich, Switzerland
\\$^2$Institute of Astronomy, Cambridge University, Madingley Road, 
Cambridge, CB3 OHA, England
\\$^3$INAF - Osservatorio Astronomico di Trieste, Via G.B. Tiepolo 11, 
I-34131 Trieste, Italy
}

\begin{document}

\maketitle

\begin{abstract}
Using analytic arguments and a suite of very high resolution ($\sim 10^3$\,M$_\odot$ per particle) cosmological hydro-dynamical simulations, we argue that high redshift, $z \sim 10$, $M \sim 10^8$\,M$_\odot$ halos, form the smallest `baryonic building block' (BBB) for galaxy formation. These halos are just massive enough to efficiently form stars through atomic line cooling and to hold onto their gas in the presence of supernovae winds and reionisation. These combined effects, in particular that of the supernovae feedback, create a sharp transition: over the mass range $3-10 \times 10^7$M$_\odot$, the BBBs drop two orders of magnitude in stellar mass. Below $\sim 2 \times 10^7$M$_\odot$ galaxies will be dark with almost no stars and no gas. Above this scale is the smallest unit of galaxy formation: the BBB.

We show that the BBBs have stellar distributions which are spheroidal, of low rotational velocity, old and metal poor:  they resemble the dwarf spheroidal galaxies (dSphs) of the Local Group (LG). Unlike the LG dSphs, however, they contain significant gas fractions. We connect these high redshift BBBs to the smallest dwarf galaxies observed at $z=0$ using linear theory. A small fraction ($\sim 100$) of these gas rich BBBs at high redshift fall in to a galaxy the size of the Milky Way. We suggest that ten percent of these survive to become the observed LG dwarf galaxies at the present epoch. This is consistent with recent numerical estimates. Those in-falling halos on benign orbits which keep them far away from the Milky Way or Andromeda manage to retain their gas and slowly form stars - these become the smallest dwarf irregular galaxies; those on more severe orbits lose their gas faster than they can form stars and become the dwarf spheroidals. The remaining 90\% of the BBBs will be accreted. We show that this gives a metallicity and total stellar mass consistent with the Milky Way old stellar halo.
\end{abstract}


\section{Introduction}\label{sec:introduction}

The Local Group (LG) of galaxies provide a unique test-bed for galaxy formation theories and cosmology. Their close proximity allows individual stars to be resolved giving accurate kinematics, stellar populations and star formation histories (see e.g. \bcite{2001ApJ...563L.115K} and \bcite{2002MNRAS.332...91D}); their spatial distribution can be compared with cosmological predictions to give useful constraints \citep{2005astro.ph.10370M}; and their large mass to light ratios can be used, through dynamical modelling, to place constraints on the nature of dark matter \citep{2001ApJ...563L.115K}.

The LG dwarf galaxies are usually split into
three types: dSph galaxies, which typically have old stellar
populations, are spheroidal 
in morphology, lie close to their host galaxy\footnote{We use the
  terminology `host galaxy' throughout this paper to refer to either the
  Milky Way (MW) or Andromeda (M31) depending on which of these is closer to the
  satellite being discussed.} and are devoid of HI
gas; dIrr galaxies, which have younger stellar populations, irregular
morphology, lie further away from their host galaxy and contain
significant HI gas; and the transition galaxies, which are in between
the dSph and dIrr types \citep{1998ARA&A..36..435M}.

In our current `vanilla' cosmological paradigm ($\Lambda$CDM - cold dark matter with a cosmological constant), all structure forms from the successive mergers of smaller substructures \citep{1978MNRAS.183..341W}. While this theory has been tremendously successful on scales larger than $\sim 1$\,Mpc, on smaller scales it has fared less well (see e.g. \bcite{2004ApJ...612..628D}). A now long-standing puzzle is the `missing satellites' problem: there appears to be an order of magnitude fewer satellite galaxies in the LG than would naively be predicted from the mass function of dark matter halos (see e.g. \bcite{1993MNRAS.264..201K} and \bcite{1999ApJ...522...82K}). A number of solutions to this puzzle have been presented, the two main threads being either to alter the nature of dark matter (\bcite{2001ApJ...556...93B} and \bcite{2001ApJ...559..516A}), or to invoke some form of feedback from supernovae explosions (see e.g. \bcite{1974MNRAS.169..229L} and \bcite{2000MNRAS.317..697E}), or photo-evaporation (see e.g. \bcite{1996MNRAS.278L..49Q} and \bcite{1999ApJ...523...54B}). In the feedback scenario, one might naively expect only the most massive substructure satellites, with the deepest potential wells, to form stars and remain visible in the LG at the present epoch \citep{2002MNRAS.335L..84S}. However, this presents a problem since the most massive substructure dark matter halos predicted by $\Lambda$CDM models have central stellar velocity dispersions which are factors of 2-3 larger than those observed in the LG satellites, even after extreme tidal stripping of shocking of these halos (\bcite{2003ApJ...584..541H}, \bcite{2004ApJ...608..663K} and 
\bcite{2006MNRAS.tmp..153R}). 

An alternative view, is that the LG satellites are fossil galaxies left over from reionisation\footnote{The epoch of reionisation is caused by UV flux emitted by the first forming massive stars. Throughout this paper we suggest that reionisation occured at $z \sim 10$. Observationally there are two bounds on this epoch. The old WMAP satellite data favour larger redshifts \citep{2003ApJS..148..175S}; the data from Quasar absorption spectra favour smaller redshifts (\bcite{2005astro.ph.12082F}; but see also \bcite{2006NewAR..50...94B}); the new WMAP data favour our chosen redshift \citep{2006astro.ph..3449S}.} (\bcite{2000ApJ...539..517B}, \bcite{2001ApJ...548...33B}, \bcite{2002MNRAS.333..177B}, \bcite{2004ApJ...609..482K}, \bcite{2004ApJ...600....1S}, \bcite{2005astro.ph..9402K}, \bcite{2005ApJ...629..259R} and \bcite{2006astro.ph..1401G}). In this scenario, only those rare over-dense peaks which collapse before redshift $z \sim 10$ (the epoch of reionisation) and achieve a potential well deep enough to form stars remain visible at the present epoch; the remaining satellite galaxies have star formation quenched by the background UV flux from reionisation. In this model it is not the most massive substructure halos at $z=0$ which are the LG satellites, but rather the {\it survivors} from halos which form stars at $z \sim 10$. There is observational evidence for such a scenario, both indirectly from high redshift quasar absorption spectra \citep{Wyithe:2006st}, and from the star formation histories of LG dIrrs,  which show a strong suppression in star formation up to $z \sim 1$ \citep{2005NewAR..49..453S}.

While this general model has been investigated by a number of authors in the literature, there are significant differences in the details. \citet{2004ApJ...609..482K} argue that only a few of the LG satellites are genuine fossils; the majority were significantly more massive in the past, formed most of their stars after reionisation (at $z \sim$ 3), and then subsequently lost their mass through tidal stripping and shocking. They find that the central velocity dispersions can be sufficiently lowered in their model once satellite-satellite interactions are taken into account alongside stripping and shocking from the Milky Way (MW) or Andromeda (M31). This model complements the `tidal model' proposed by  \citet{2001ApJ...559..754M} and \citet{2001ApJ...547L.123M} to explain the distance-morphology relation between dSphs and dIrrs. In these two papers, it is suggested that  all of the Local Group satellite galaxies started out looking more like the dIrrs with a disc-like morphology. Those satellites on orbits which brought them close to the MW or M31 then formed induced bars which buckled leaving a spheroidal remnant: a dSph.

\citet{2005ApJ...629..259R} and \citet{2006astro.ph..1401G} focus on the pure fossils left over from reionisation. They present a detailed cosmological model which includes radiative transfer and molecular cooling from H$_2$. These new key ingredients allow them to study star formation in mini halos with virial temperatures $T < 10^4$\,K - the temperature at which hydrogen starts to become collisionally ionised. They find, contrary to previous studies, that such mini-halos can cool efficiently and form stars. They posit that such halos could then (if they survive) be the progenitors of dSph galaxies in the LG. They require more massive halos to then become the dIrr galaxies. However, they do naturally recover the spheroidal morphology, low gas fractions, low rotational velocity and old stellar populations observed in the LG dSphs (see e.g. \bcite{1998ARA&A..36..435M}), without recourse to any tidal transformations. 

\citet{2005ApJ...629..259R} and \citet{2006astro.ph..1401G} study in detail the effects of photo-ionising feedback from star formation, but they do not include the effects of feedback from supernovae winds. It is important to separate feedback from supernovae (which is included in the models by \citet{2005ApJ...629..259R} and \citet{2006astro.ph..1401G}) from feedback from supernovae {\it winds}, which are not. In this paper, we implement and test the effect of both kinds of feedback. This is a key difference in the work we present here. Observationally, we can see that supernovae winds are driven in galaxies undergoing a phase of star formation and that such winds are important, especially on the scale of dwarf galaxies (see e.g. \bcite{2005MNRAS.358.1453O}). Theoretically, \citet{1999ApJ...513..142M} and \citet{Marcolini:2006mk} have shown that there is a transition scale at about $M_{\mathrm{crit}} \sim 10^7 - 10^8$\,M$_\odot$ below which dwarf galaxies efficiently lose their gas from supernovae winds\footnote{We note, however, that this conclusion is degenerate with the mechanical luminosity in the wind. For the mean plausible range, efficient mass loss occurs at $\sim 10^7$M$_\odot$.}; while \citet{1986ApJ...303...39D} and \citet{2003Astro-Ph..0210454} demonstrated that supernovae feedback could account for the global scaling relations of the LG dwarfs. 

In this paper, we use a suite of very high resolution cosmological hydro-dynamical simulations, which include gas cooling, star formation, feedback from supernovae, galactic winds and reionisation, to study a new model for the formation of the LG dwarf galaxies. We do not include the effects of the detailed radiative transfer and cooling physics required to model star formation in mini-halos with mass $M \simlt 10^7$\,M$_\odot$, since we are interested primarily in halos more massive than this. We suggest that the smallest dwarf galaxies of the LG share a common progenitor: rare, $\sim 3 \sigma$, $\simgt 10^8$\,M$_\odot$, dark matter halos at $z \sim 10$. These halos are just massive enough to efficiently form stars through atomic line cooling and to hold onto their gas in the presence of supernovae winds and reionisation. As a result they are the smallest  `baryonic building block' (BBB) available for galaxy formation. 

Some of these gas-rich early forming galaxies fall-in late to the LG and survive as dwarf galaxies. Those in-falling halos on benign orbits which keep them far away from the MW or M31 manage to retain their gas and slowly form stars - these become the dIrrs; those on more severe orbits lose their gas faster than they can form stars and become the dSphs. This suggests that the dIrrs should also have an old extended spheroidal component of stars: a stellar halo. There is increasing observational evidence that this is indeed the case (\bcite{1996ApJ...467L..13M}, \bcite{2003Sci...301.1508M} and \bcite{2000AJ....119..177A}; and see \bcite{2005AJ....130.1593V} for a study of dIrrs outside of the LG). It also suggests that the star formation histories of the LG dwarfs should all show an early, pre-reionisation burst of star formation. This also appears to be the case (see e.g. \bcite{2000MNRAS.317..831H} and \bcite{2002MNRAS.332...91D}).

The idea of a smallest building block has a long history in the literature. \citet{1953ApJ...118..513H} pointed out that gas cooling becomes efficient for ionised hydrogen at $10^4$\,K; \citet{1968ApJ...154..891P} presented a model with a smallest mass block of $10^5$M$_\odot$ for star formation in the context of globular clusters; \citet{1984ApJ...277..470P} updated this argument to include dark matter and showed that the relevant mass was $\sim 10^8$M$_\odot$, as suggested here; and, from an observational point of view, \citet{1978ApJ...225..357S} showed that the galactic globular cluster abundances are not correlated with distance -- a result which has led to the hierarchical merging model being currently favoured over the then-popular model of monolithic collapse \citep{1962ApJ...136..748E}. Here, we take the next logical step by investigating the morphology and kinematics of the stars, gas and dark matter in these high redshift BBBs, and making a link to the smallest galaxies observed in the Local Group at the present epoch. 

To test our model, we use a small box size of 1\,Mpc and stop the simulation at $z=10$ (to avoid simulations becoming non-linear on the scale of the box). With such a small box size we achieve an unprecedented mass resolution of $\sim 10^3$\,M$_\odot$ per particle. This allows us to accurately track the kinematics and morphology of galaxies of total mass $M \sim 10^8$\,M$_\odot$ and compare our results with observations from the LG dwarfs. Ideally, one would like to use a larger box and evolve all the way to redshift $z=0$. However, this is not technically feasible for a mass resolution of $\sim 10^3$\,M$_\odot$ per particle, at the present time. Instead we have to compromise by stopping at redshift $z=10$ and making the link to redshift $z=0$ using a mixture of linear theory arguments (section \ref{sec:motivation}) and results from other studies in the literature (\bcite{2005astro.ph.10370M} and \bcite{2005astro.ph..4277M}).

This paper is organised as follows: in section \ref{sec:motivation} we present some theoretical motivation for our model. In section \ref{sec:hydro} we describe the simulations. We used a control simulation with only dark matter, and five other simulations which explore the effect of the star formation prescription and supernovae winds of varying strength. All simulations were run with identical initial phase space distributions. In section \ref{sec:results} we describe the results from our suite of high resolution simulations. In section \ref{sec:discussion}, we discuss our results in the context of recent observations. Finally, in section \ref{sec:conclusions}, we present our conclusions. 

\begin{figure}
\begin{center}
\includegraphics[width=8.25cm]{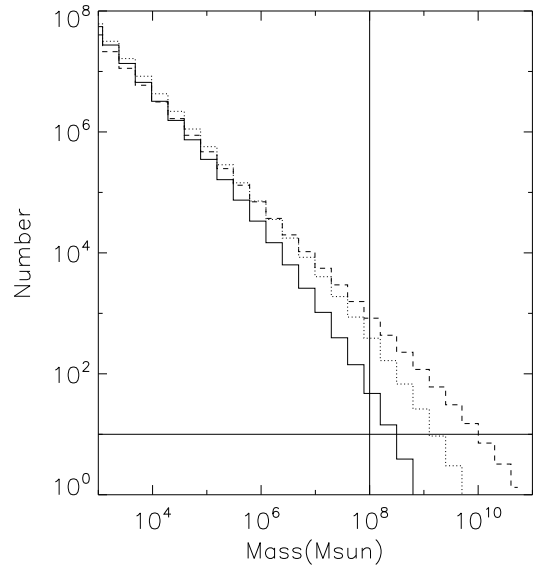}
\caption{The number of halos, $N$, formed up to redshift $z = 15,10,5$ (solid, dotted and dashed lines), in a given mass range which are likely to have fallen into a larger halo of mass $M_0 = 2\times 10^{12}$\,M$_\odot$ at $z_0=0$. Over-plotted are the number of dSph galaxies currently observed around the MW (horizontal line) and the mass scale $M_\mathrm{crit}$ (vertical line).}
\label{fig:numberhalo}
\end{center}
\end{figure}

\section{Theoretical motivation: the `baryonic building block'}\label{sec:motivation}

The mass scale, $M_{\mathrm{crit}} \sim 10^8$\,M$_\odot$, is interesting. Firstly, it corresponds to the mass below which a galaxy cannot retain its gas in the presence of supernovae winds (\bcite{1999ApJ...513..142M}; \bcite{Marcolini:2006mk}\footnotemark). Secondly, the virial temperature of halos at a given redshift, $T_\mathrm{v}$, is given by \citep{1999coph.book.....P}:

\begin{equation}
T_\mathrm{v}/K = 10^{5.1}(M/10^{12}\mathrm{M}_\odot)^{2/3}(f_c \Omega h^2)^{1/3}(1+z)
\label{eqn:virialt}
\end{equation}
\noindent
where $M$ is the mass of the halo; $z$ is the redshift; $h = 0.72$ is the dimensionless Hubble parameter at the present epoch; $\Omega = \Omega(z)$ is the ratio of the halo density to the critical density; and $f_c$ is the density enhancement of the collapsing halo with respect to the background. Working at high redshift ($z=10$) has the advantage that $\Omega \simeq 1$ irrespective of the cosmological model. Spherical top-hat collapse then gives $f_c \simeq 178$ at virialisation \citep{1999coph.book.....P}. 

\footnotetext{In fact \citet{Marcolini:2006mk} find that $M_\mathrm{crit} \sim 2\times 10^7$M$_\odot$ as a result of metal cooling. This slightly lower bound doesn't affect our argument.}

The temperature, $T_\mathrm{v} = 10^4$\,K, is the temperature scale at which hydrogen starts to collisionally ionise. This allows for efficient atomic line cooling. Below this temperature, hydrogen can only cool through radiative emission from H$_2$ roto-vibrational transitions, which is very inefficient by comparison \citep{1999MNRAS.305..802L}. Using $T_\mathrm{v} = 10^4$\,K and $z=10$ gives $M = 6.4\times 10^7$\,M$_\odot \simeq M_{\mathrm{crit}}$. Thus, just before the epoch of reionisation ($z=10$), halos which reach a virial temperature of $T_\mathrm{v} = 10^4$\,K and can efficiently form stars are those of mass $M_\mathrm{crit}$. 

Thirdly, the Extended Press Schechter (EPS) scheme \citep{1993MNRAS.262..627L} can be used to calculate the mean number of halos of a given mass and redshift, $M,z$, which fall into a larger halo of given mass and redshift, $M_0, z_0$:

\begin{equation}
\frac{dN}{d\ln M} = \sqrt{\frac{2}{\pi}} M_0 \sigma_0(M) \frac{D}{S^{3/2}}\exp \left(-\frac{D^2}{2S}\right)\left|\frac{d\sigma_0(M)}{dM}\right|
\label{eqn:eps}
\end{equation}
\noindent
where $D = \delta_c \left[D(z)^{-1} - D(z_0)^{-1}\right]$; $S = \sigma_0^2(M) - \sigma_0^2(M_0)$; $\sigma_0^2(M)$ is the variance of the linear power spectrum at $z=0$ smoothed with a top-hat filter of mass $M$; $\delta_c$ is the critical over-density for spherical collapse; and $D(z)$ is the growth factor which depends on the cosmology\footnotemark. 

\footnotetext{Usually it is stated that $\delta_c = 1.69$, but this is only true for a universe with $\Omega_\mathrm{m} = 1$; where $\Omega_\mathrm{m}$ is the ratio of the matter density of the universe to the critical density required for closure. Here we calculate $\delta_c$ and $D(z)$ correctly (numerically) for a $\Lambda$CDM cosmology in which $\Omega_\mathrm{m}+\Omega_\Lambda = 1$; where $\Omega_\Lambda$ is the ratio of the dark energy density to the critical density, or the `cosmological constant'. The relevant equations for this are given in \citet{1999ApJ...511....5E}. The cosmological parameters we use are given in section \ref{sec:hydro}.}

Integrating equation \ref{eqn:eps} allows us to solve for the number of halos, $N$, formed from the beginning of the universe {\it up to} some redshift $z$, in a given mass range, which are likely to have fallen into a larger halo of mass $M_0 = 2\times 10^{12}$\,M$_\odot$ ($\sim$ the mass of the MW \citep{1999MNRAS.310..645W}) at $z_0=0$. This is what we plot in Figure \ref{fig:numberhalo} for $z=15,10,5$ (solid, dotted and dashed lines). The formation redshift for these halos, $z$, then corresponds to the epoch of reionisation, before which the LG dwarfs spheroidals could form stars. Over-plotted are the number of dSph galaxies currently observed around the MW (horizontal line) and the mass scale $M_\mathrm{crit}$ (vertical line). For early reionisation $z \simgt 10$, such as that favoured by the WMAP data \citep{2003ApJS..148..175S}, the total number of satellites with mass $10^7 - 10^8$\,M$_\odot$ is of order 100. This is an order of magnitude greater than the observed number in the Local Group. However, it is the number which will be {\it accreted} by a MW sized halo, not the number which survive as satellite halos at $z=0$. \citet{2005astro.ph.10370M} have recently shown that only $\sim 10$\% survive, giving the correct order of magnitude of satellites around the MW. A final encouraging point is that the total stellar mass in the accreted satellites is then $\sim 10^8$M$_\odot$, which is the mass of the MW old stellar halo \citep{1998gaas.book.....B}. Note that the above requires early reionisation. If reionisation is found to occur much later ($z \sim 5$), this would rule out our model. Turning this around, we tentatively suggest that the LG dwarf galaxies' number and distribution can constrain the epoch of reionisation. 

$M_\mathrm{crit} \sim 10^8$\,M$_\odot$ appears to be a critical mass scale at which gas cooling becomes efficient so that star formation can occur, but also at which the potential well is just deep enough to hold on to the remaining gas left over from star formation in the presence of supernovae winds. The number of surviving subhalos of this mass formed at $z \sim 10$ which fall into a MW sized halo at $z=0$ is consistent with the observed number of LG satellites; while observations of the LG and nearby dwarfs suggest masses of the order $10^8$\,M$_\odot$ (see Figure \ref{fig:mtol}). These facts, combined with the mounting evidence for old stellar halos in the dIrr galaxies, and star formation histories which show pre-reionisation bursts, form the main motivation for investigating our model in more detail.

\begin{figure*}
\begin{center}
\includegraphics[width=5.8cm]{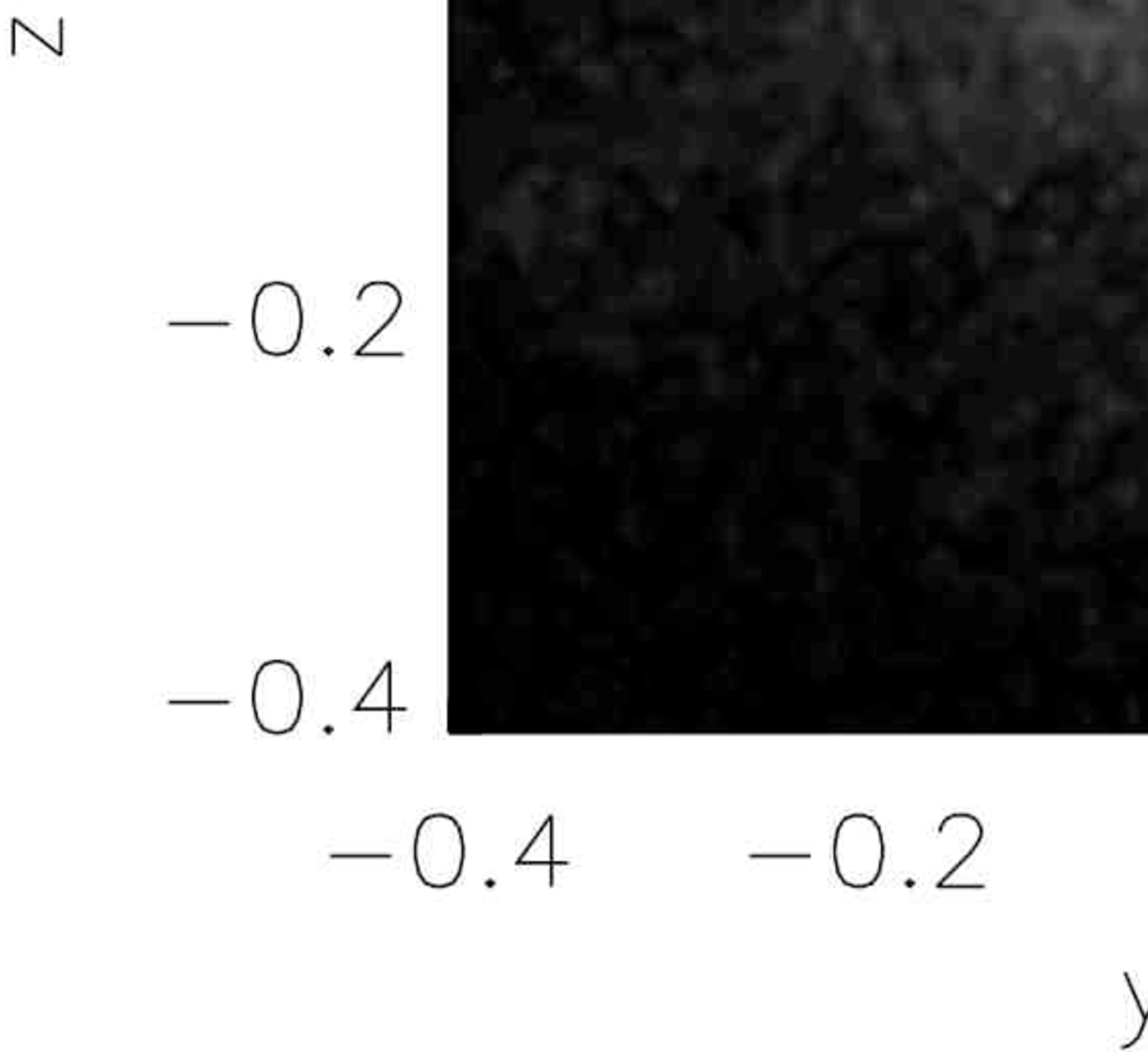}
\includegraphics[width=5.8cm]{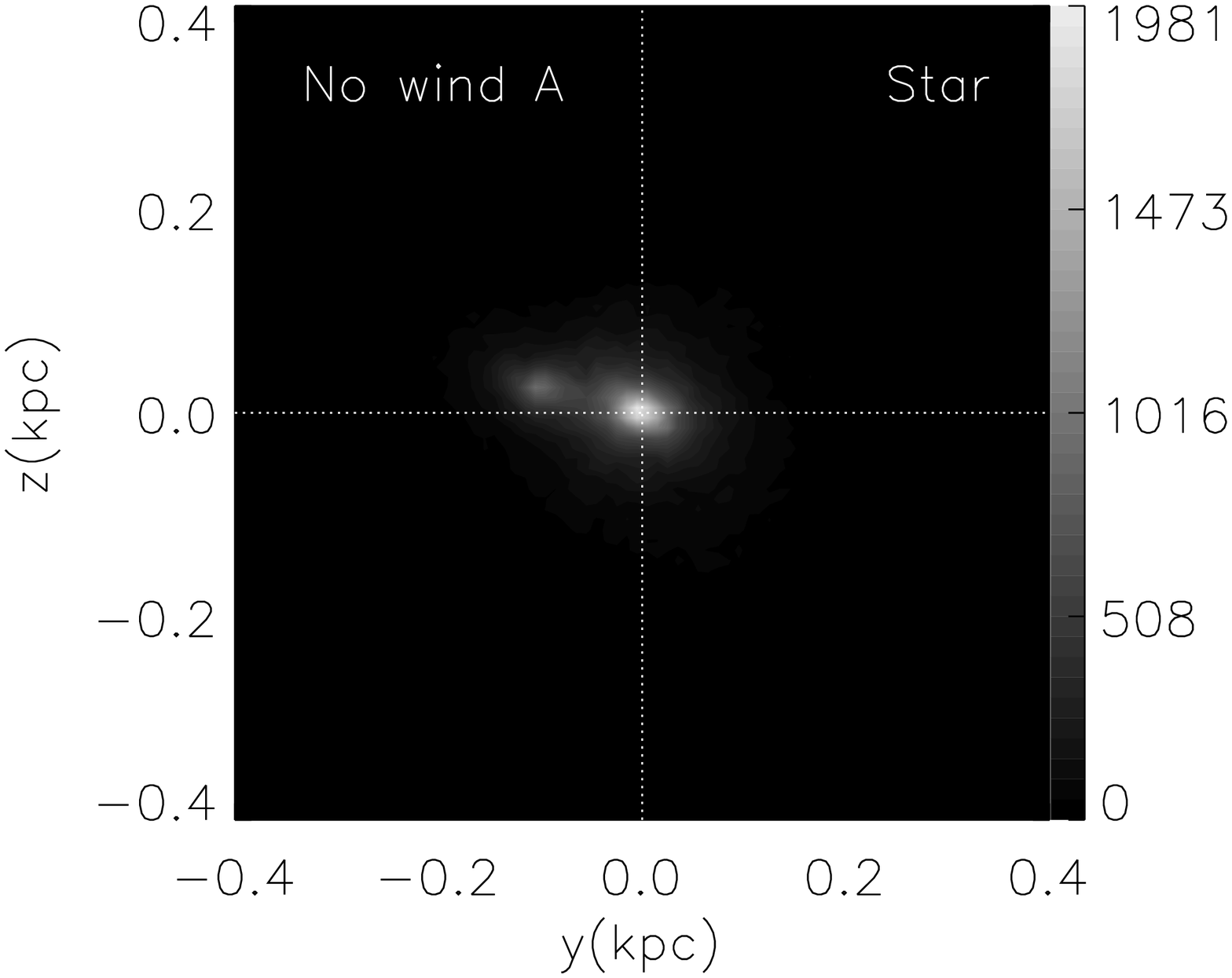}
\includegraphics[width=5.8cm]{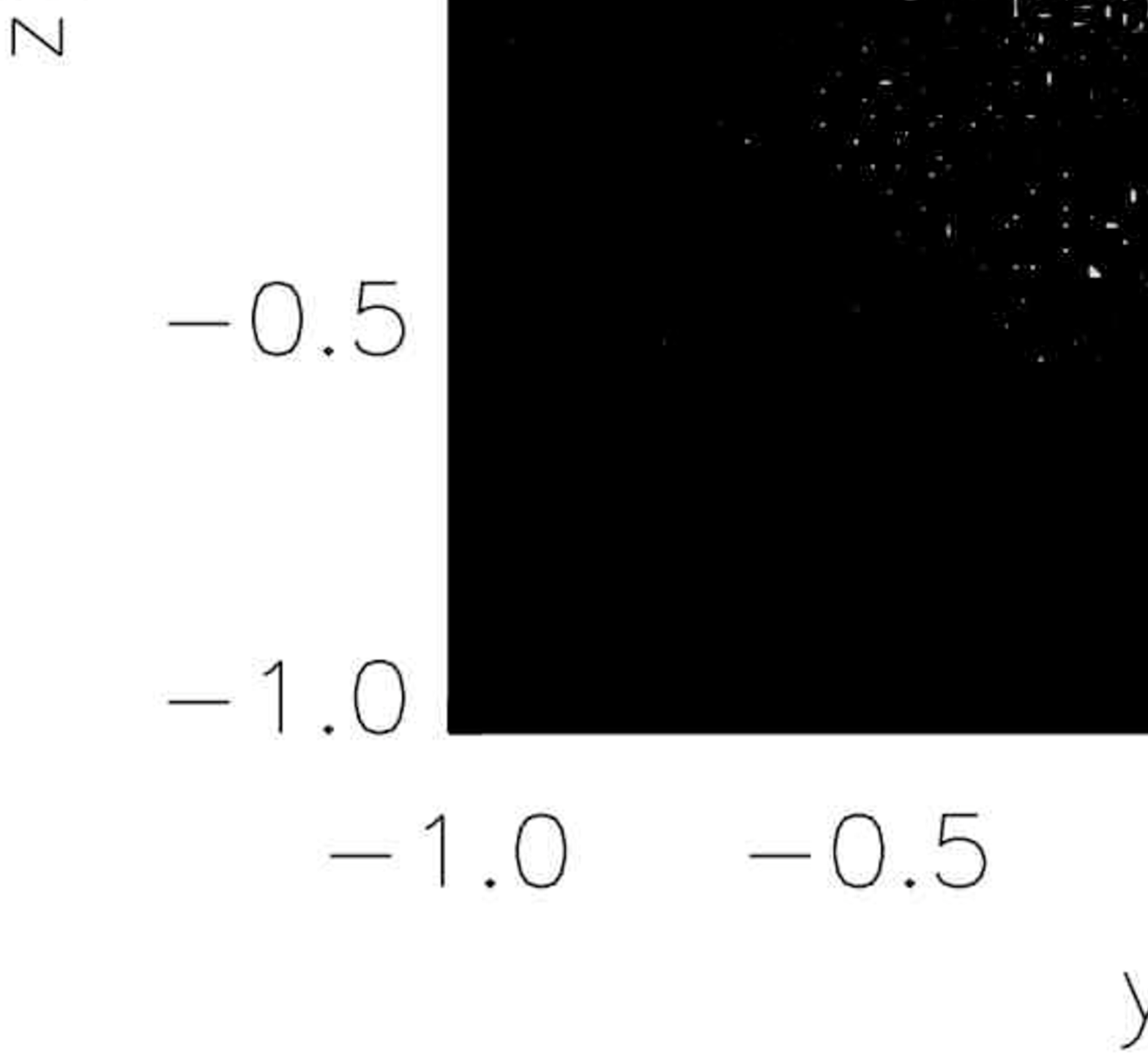}\\
\vspace{-10mm}
\includegraphics[width=5.8cm]{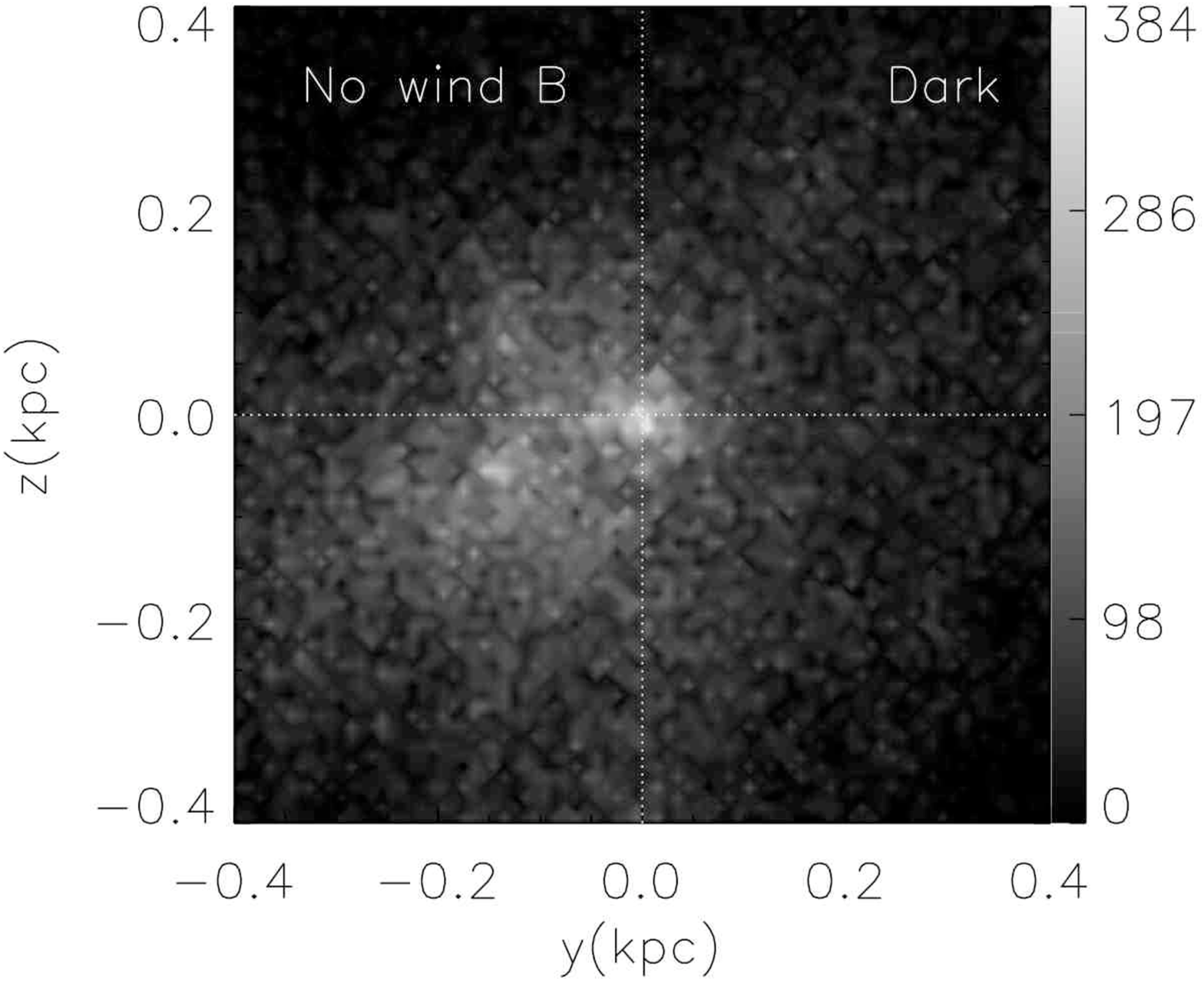}
\includegraphics[width=5.8cm]{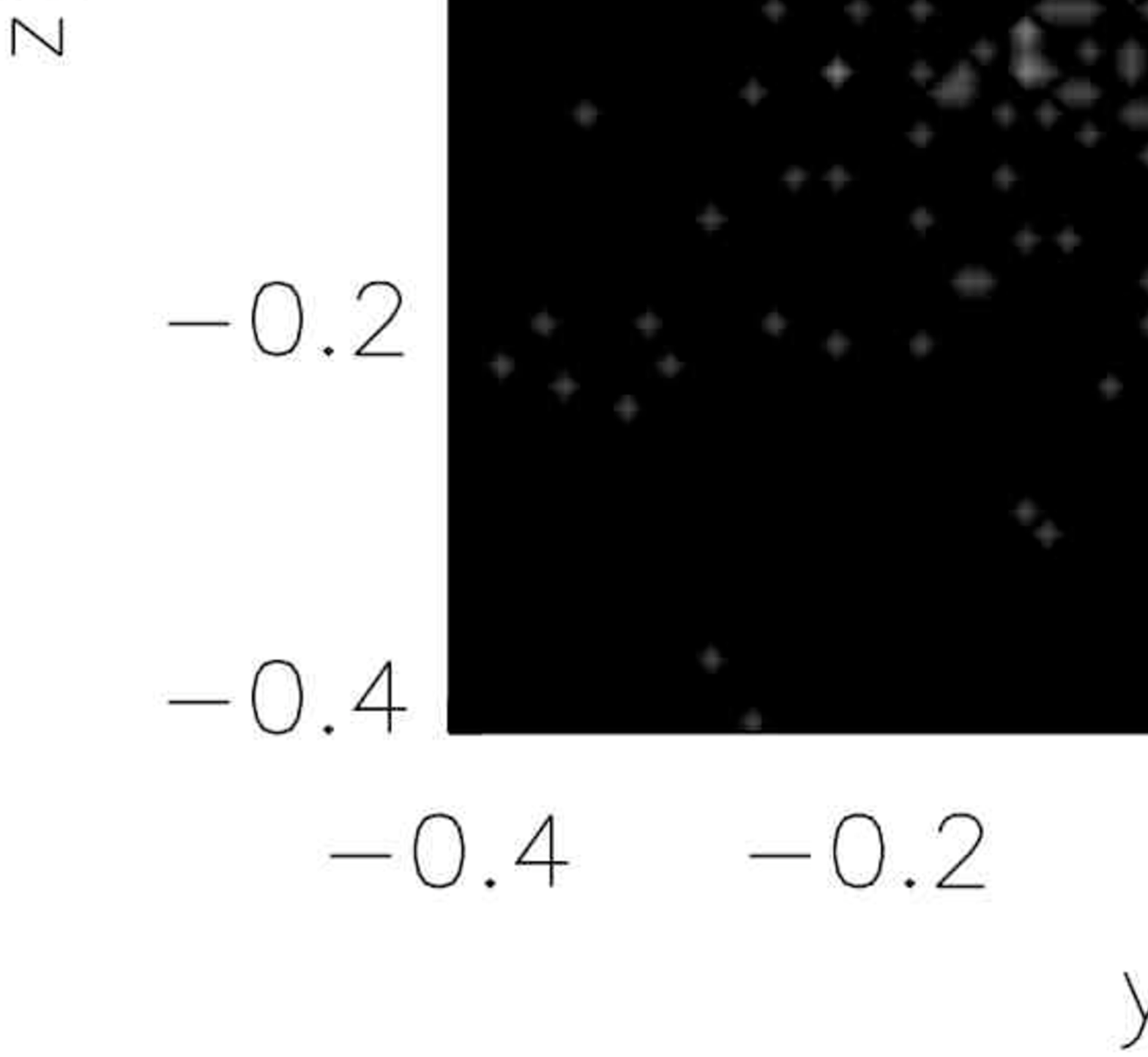}
\includegraphics[width=5.8cm]{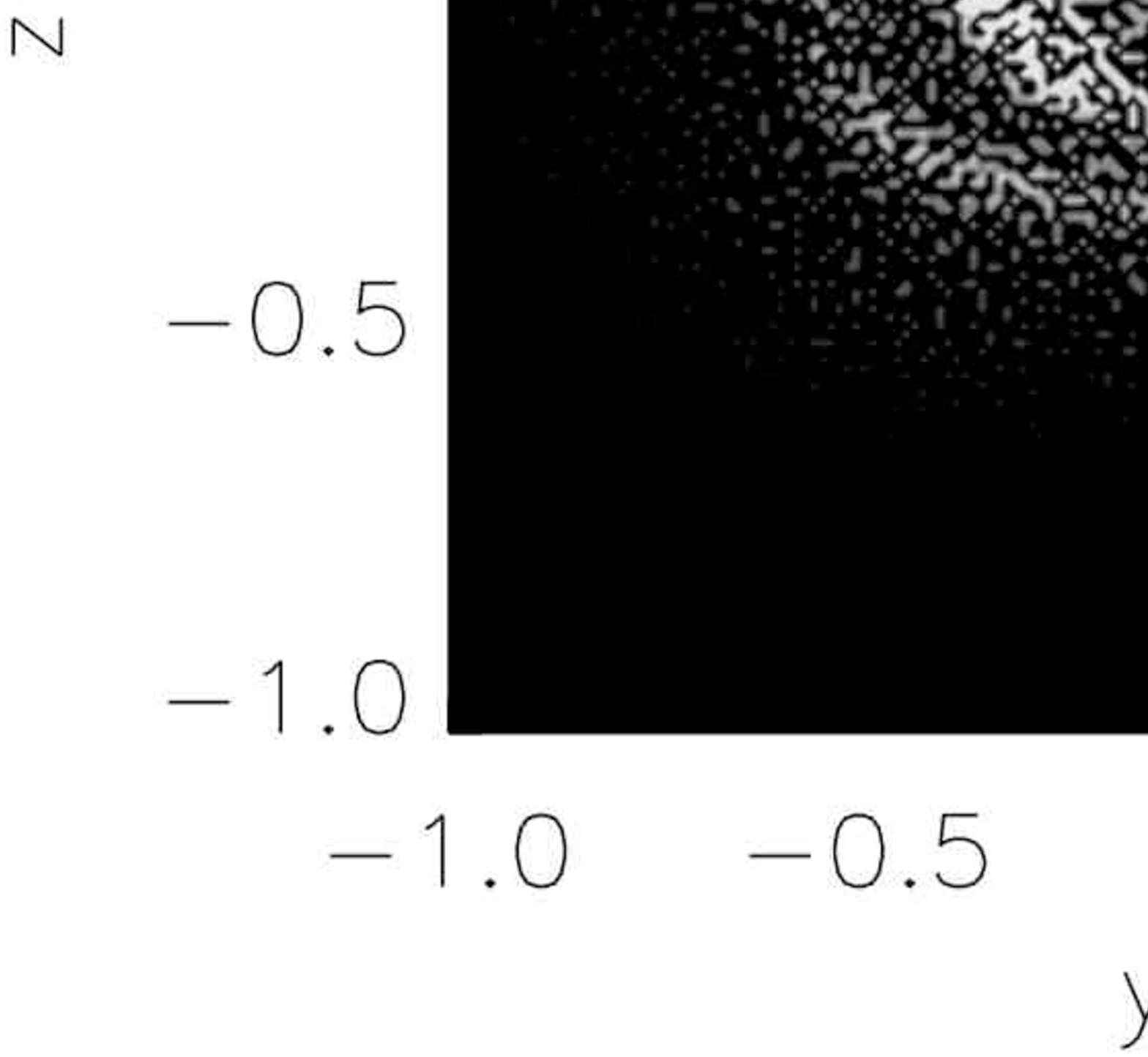}\\
\vspace{-10mm}
\includegraphics[width=5.8cm]{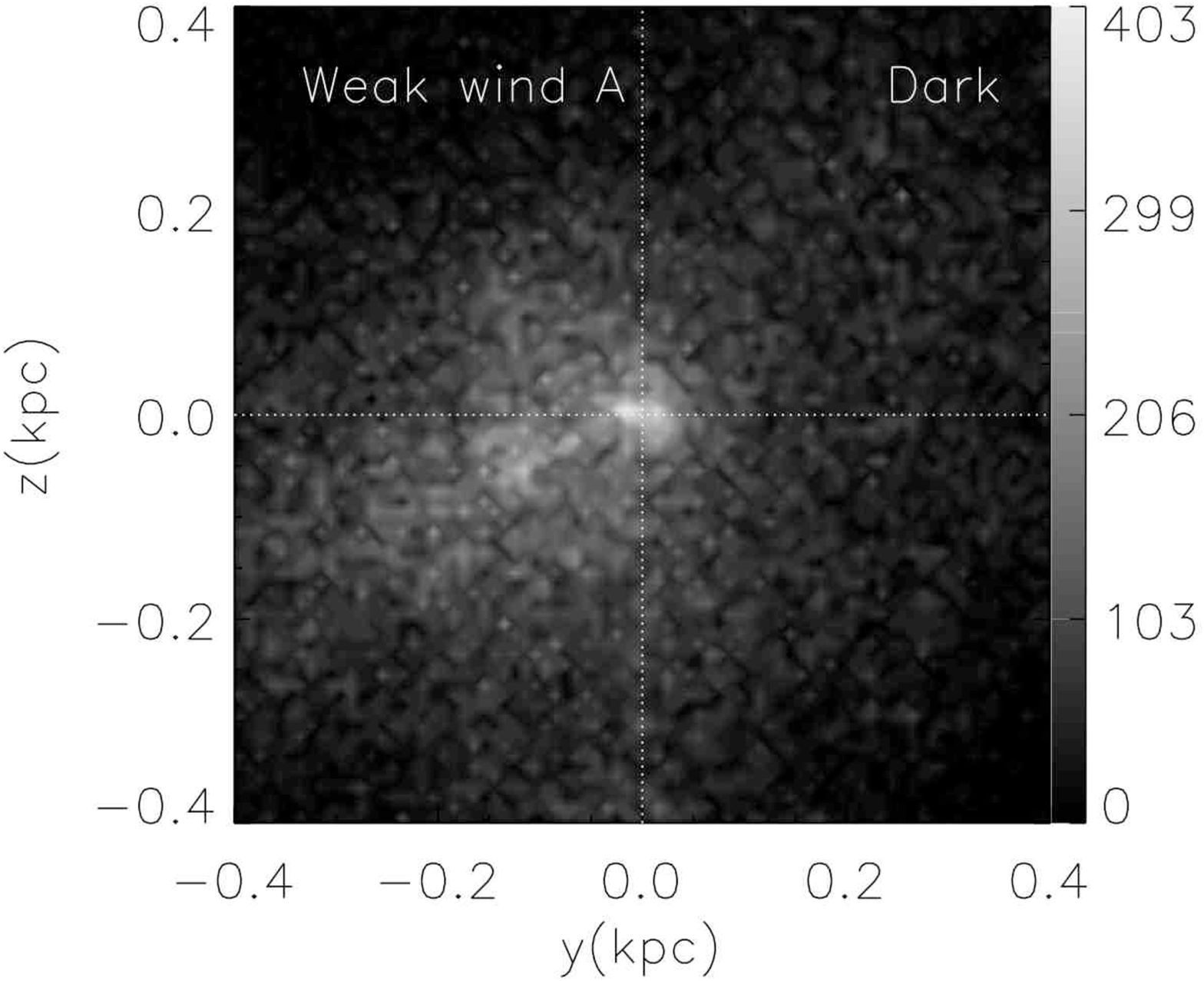}
\includegraphics[width=5.8cm]{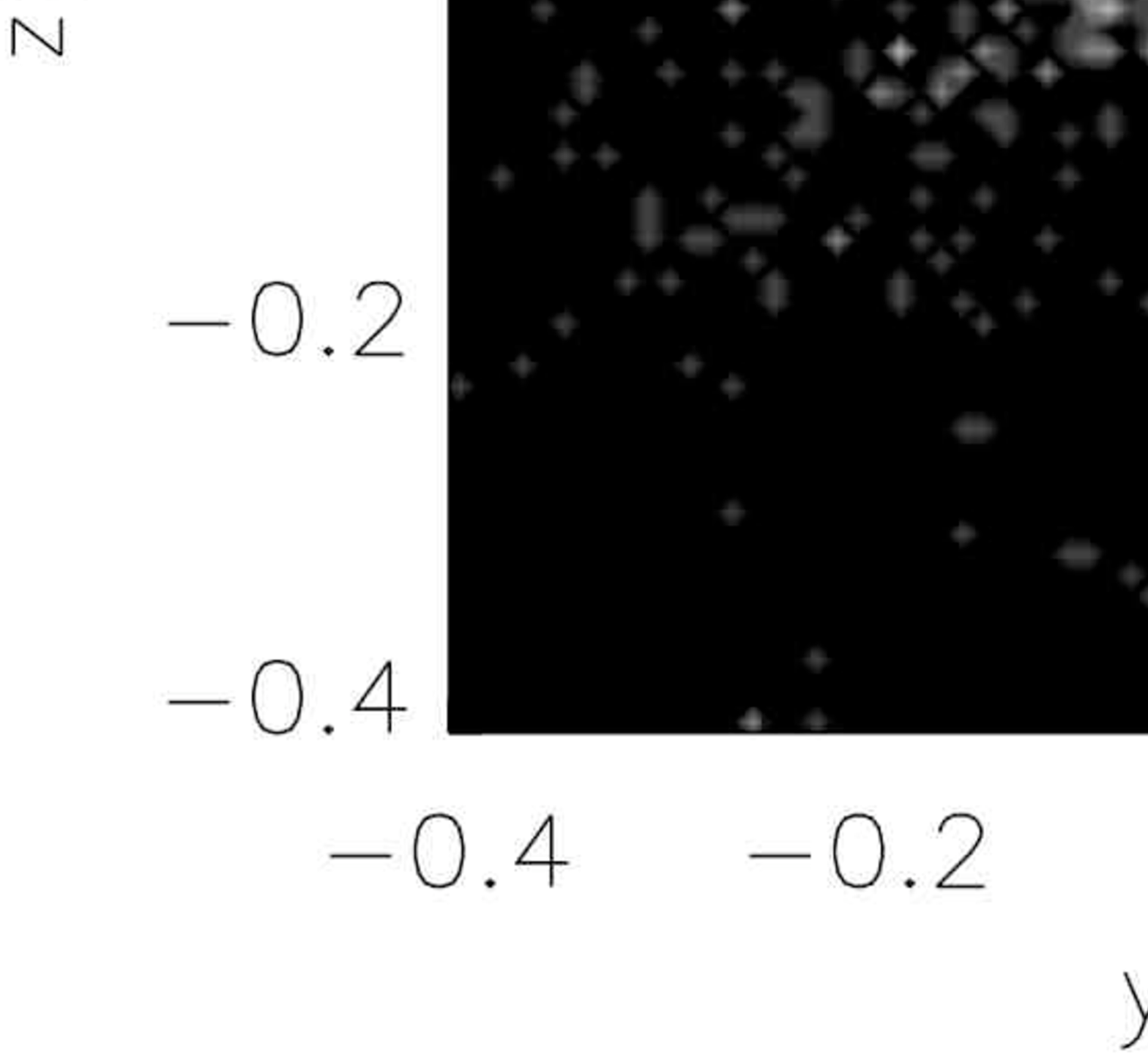}
\includegraphics[width=5.8cm]{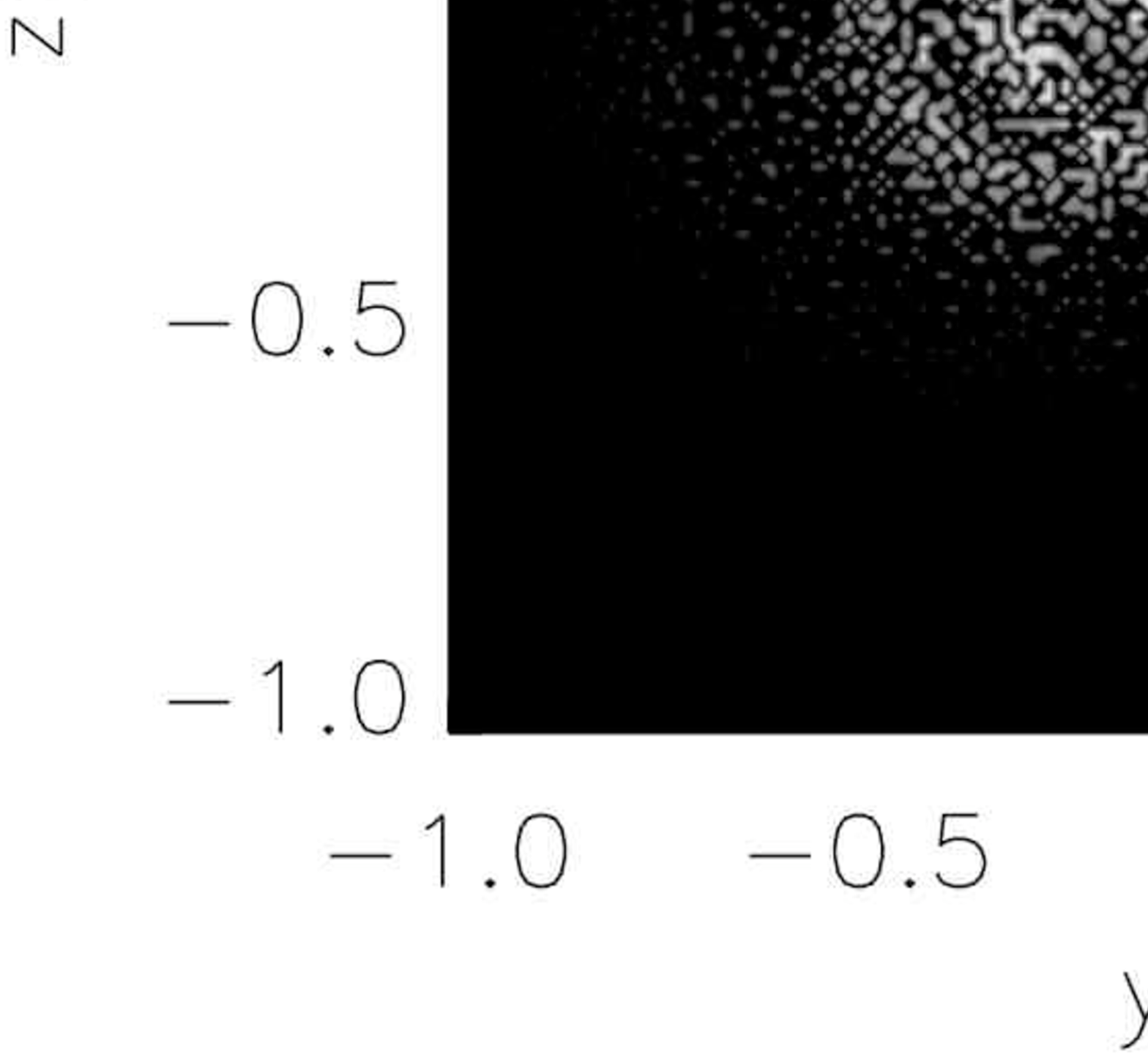}\\
\vspace{-10mm}
\includegraphics[width=5.8cm]{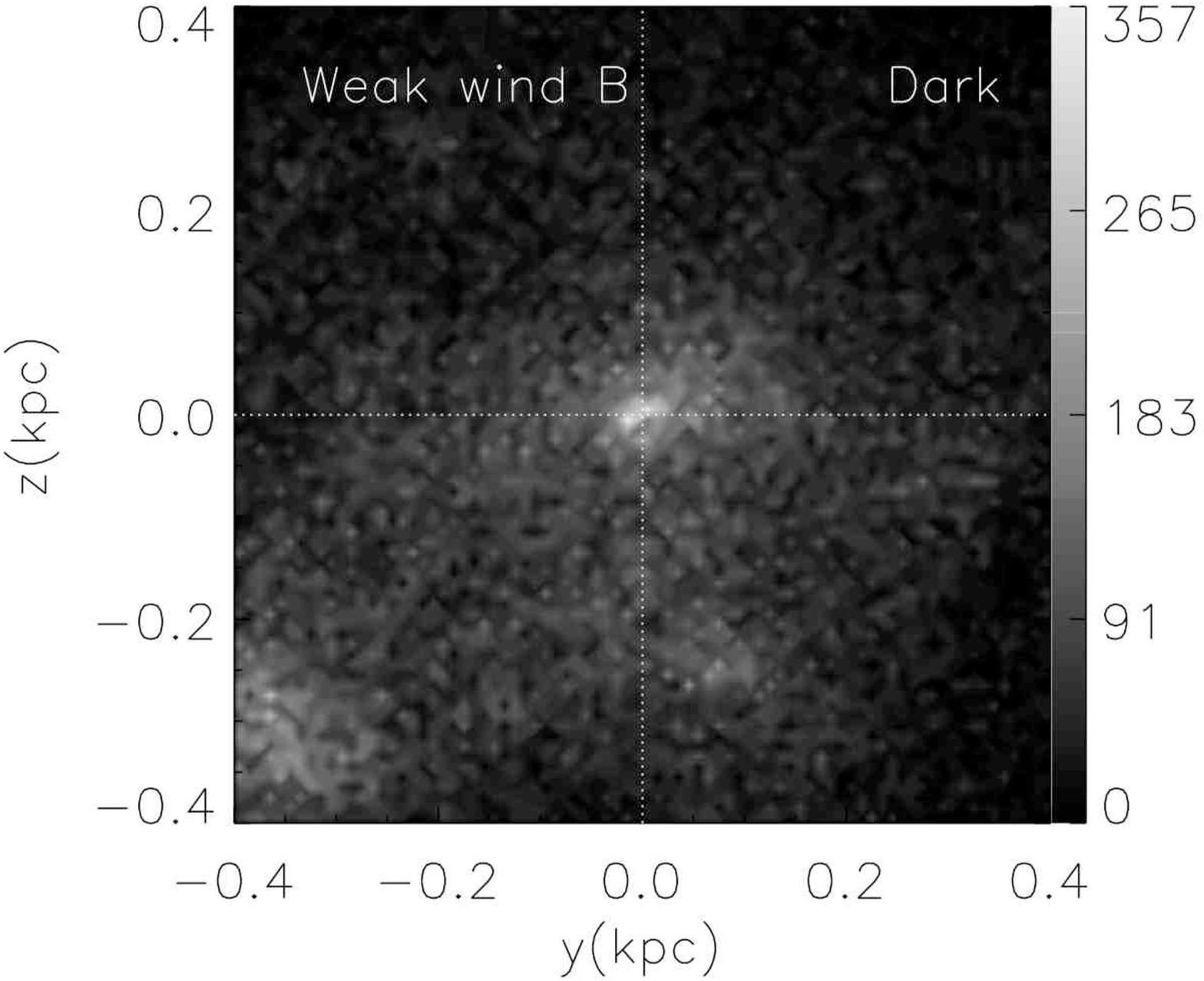}
\includegraphics[width=5.8cm]{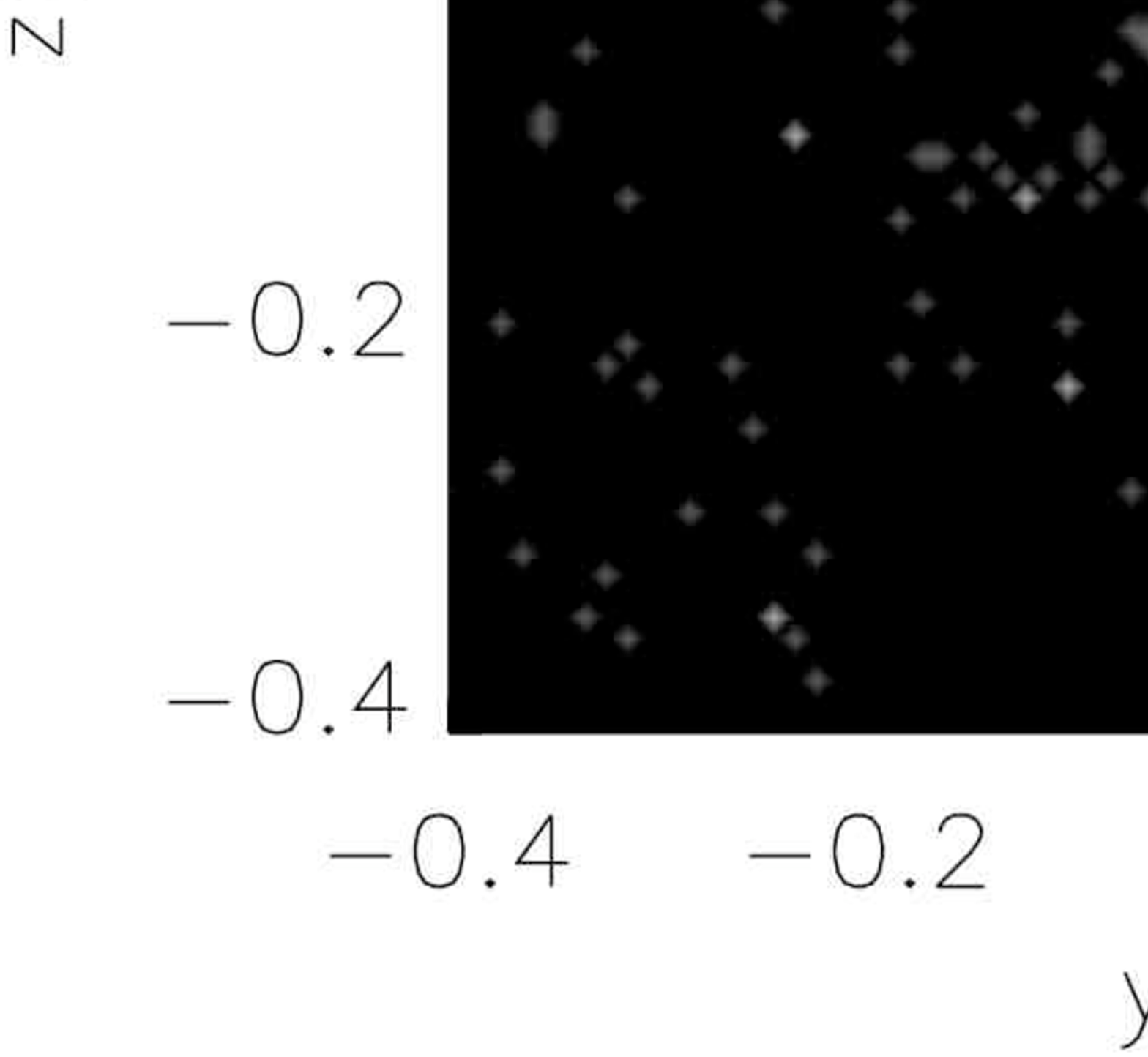}
\includegraphics[width=5.8cm]{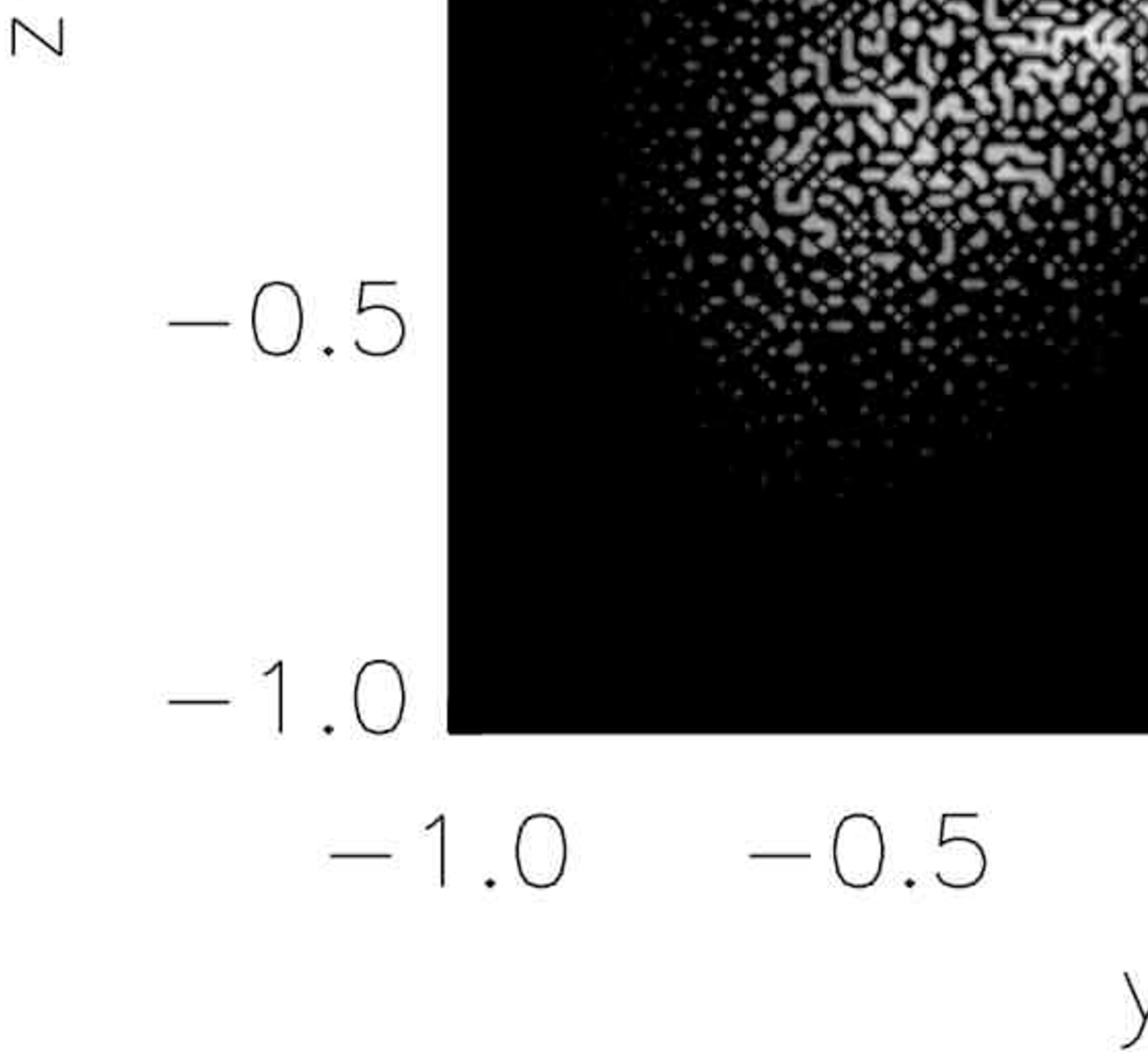}
\vspace{-10mm}
\caption{Projected density for the dark matter (left), stars (middle) and gas (right), for (from top to bottom) the NWA, NWB, WWA and WWB simulations; second most massive halo. The gas density is for a thin slice of width 0.1kpc about the x-axis. The contour bars show density in units of $10^6$\,M$_\odot$kpc$^{-3}$ in all cases except the NW gas density, which is in units of $10^4$\,M$_\odot$kpc$^{-3}$. Note that the gas is more extended and less concentrated than the stars and dark matter and for this reason is plotted on a different scale. All simulations were run with identical initial phase space distributions. This means the above halos are identical apart from their star formation and feedback prescriptions. The key thing to notice is that without any supernovae feedback (NWA) there is significant star formation; all of the most massive substructures and the central halo have bound stars. By contrast, with supernovae feedback, the NWB, WWA and WWB simulations are similar: there is very little star formation and significant remaining gas. The form of the feedback is not critical. NWB uses a heating term for the supernovae feedback; WWA uses a supernovae driven galactic wind; WWB uses both. Since NWB and WWA are so similar, we show only results for the NWB simulation from here on.}
\label{fig:overview}
\end{center}
\end{figure*}

\section{Hydro-dynamical simulations}\label{sec:hydro}

We ran hydro-dynamical simulations using
the new version of the parallel TreeSPH code {\small GADGET-2}
(\bcite{2001NewA....6...79S}; \bcite{2005MNRAS.364.1105S}).  {\small GADGET-2}
was used in its TreePM mode which speeds up the calculation of
long-range gravitational forces considerably. The simulations were
performed with periodic boundary conditions with an equal number of
dark matter and gas particles and used the conservative
`entropy-formulation' of SPH proposed by \citet{2002MNRAS.333..649S}. Radiative cooling and heating processes were followed using an
implementation similar to that of \citet{1996ApJS..105...19K} for a primordial
mix of hydrogen and helium. We assumed a mean UV background
produced by quasars as given by \citet{1996ApJ...461...20H}, and we switched
the background on at at high redshift in order to have HI reionisation
at $z\sim 13$. Such a treatment for the reionisation is quite crude: we simply stretched
the epoch of reionisation and switched on a uniform background at $z \sim13$,
instead of $z \sim 6$. However, even if the details are incorrect (to correctly follow reionisation, one would need a radiative transfer code) we obtain the behaviour we expect, i.e. a jump in the gas
temperature that brings most of the gas elements to $T\sim 10^4$\,K. Note also that the effect of reionisation only really enters at the very end of our simulations and serves only to inhibit further star formation. 

We used $2\times 448^3$ dark matter and gas particles in a 1 comoving
Mpc$/h$ box. 
The simulations were all started at $z=199$ and we have stored 19
redshift outputs for each run, mainly in the redshift range
$10<z<30$.  The initial gas temperature was $T = 546$ K, and $40\pm
2$ SPH neighbours were used to compute physical quantities. The
gravitational softening was set to 0.055 $h^{-1}$ kpc in comoving units
for all particles.

Our fiducial model was a `concordance' $\Lambda$CDM model with
$\Omega_{0{\rm m}}= 0.26$, $\Omega_{0\Lambda} = 0.74$, $\Omega_{0{\rm
b}} = 0.0463$ and $H_0=72\,{\rm km\,s^{-1}Mpc^{-1}}$, $n=1$ and
$\sigma_8=0.9$\footnotemark. This model is in agreement with most of the
observations including the parameters inferred by the WMAP team in
their first year data release and the
recent results of the Lyman-$\alpha$ forest community 
(\bcite{2003ApJS..148..175S}; \bcite{2004MNRAS.354..684V}; \bcite{2005PhRvD..71j3515S}).
The CDM transfer functions of all models have been taken from
\citet{1999ApJ...511....5E}.

\footnotetext{We use the common nomenclature where $\Omega_{0{\rm m},0\Lambda,0\rm{b}}$ are the ratio of the density of the universe in matter, dark energy and baryons respectively at redshift, $z=0$, to the critical density required for closure; $H_0$ is Hubble's constant at $z=0$; $n$ is the {\it initial} spectral index of initial matter fluctuations (1 corresponds to scale-invarience); and $\sigma_8$ is the amplitude normalisation of matter fluctuations at $z=0$: the rms density variation of the universe smoothed with a top-hat filter of radius 8 $h^{-1}$\,Mpc ($h=0.72$ is the dimensionless Hubble parameter) at $z=0$. Note that one should be cautious of this definition of $\sigma_8$ since it implicitly assumes that linear theory can connect the initial and $z=0$ amplitude of fluctuations. On scales as large as $h^{-1}$\,Mpc at $z=0$ this is usually ok, but non-linearities can certainly affect the expected peculiar velocities on these scales (see e.g. \bcite{2006MNRAS.365.1191P}).}

We note that the new three year data release by the WMAP team
\citep{2006astro.ph..3449S}
point to smaller values for the spectral index and $\sigma_8$, while
the new reionisation redshift is perfectly consistent with that chosen
in our simulations. These new values should not have a large impact on the internal structure of the halos we investigate in this paper, but the lower $\sigma_8$ will affect the statistics of such halos. We note that, when combined with data from the Lyman-$\alpha$ forest, the WMAP three year release favours higher values of $\sigma_8$, similar to those we use here (\bcite{2006astro.ph..4310V} and \bcite{2006astro.ph..4335S}). As such we felt it would be premature to re-run our simulations with lower $\sigma_8$. 

We ran six different simulations in total. One with dark matter only was run as a control simulation and is only briefly mentioned. The other five investigate different star formation prescriptions and wind strengths. In all cases we started with identical initial phase space distributions. We used two different star formation recipes. The first (A) was a very simplified prescription based only on a density and temperature cut. All gas is turned into stars when the overdensity, $\delta > 1000$ and the temperature, $T<10^5$\,K. This simple prescription does not track or calculate metallicities for the stars and gas. The second (B) was more physically motivated and modelled a sub-particle multiphase medium, including supernovae feedback. This supernovae feedback is represented by a local heating term near star forming regions and does not drive a global galactic wind. This model is described in detail in \citet{2003MNRAS.339..289S}. We used the `simplified', rather than the `explicit' mode of star formation. We list all the parameters for both prescriptions in Table \ref{tab:freeparameters}.

In addition to each star formation prescription, we investigated the effect of adding a supernovae driven galactic wind. The wind model we use is described in detail in \citet{2003MNRAS.339..289S}. We assumed that the galaxy mass-loss rate that goes into a wind, $\dot
M_{\rm w}$, is proportional to the star formation rate itself \be \dot
M_{\rm w}=\eta \dot M_\star \, , \ee where $\eta$ is a coefficient of
order unity.  Moreover, we assumed that the wind carries a fixed
fraction $\chi$ of the supernova energy. Equating the kinetic energy in
the wind with the energy input by supernovae, \be \frac{1}{2} \dot
M_{\rm w}{v_{\rm w}^2} = \chi \epsilon_{\rm SN} \dot M_\star, \ee we
obtain the wind's initial velocity as \be v_{\rm
w}=\sqrt{\frac{2 \chi \epsilon_{\rm SN}}{\eta}}, \label{eqwind} \ee with
$\epsilon_{\rm SN}= 4\times 10^{45}$ ergs/M$_\odot$, which is the average
expected value from the SN explosions' energy release ($10^{51}$ ergs).
These parametrizations and the chosen values are mainly motivated by
observations of starburst driven galactic winds in the local universe
(e.g. \bcite{1999ApJ...513..156M} and \bcite{2005MNRAS.358.1453O}).

We investigated each star formation prescription without winds (no wind A/B: NWA/B), and with winds of varying strength (weak wind A/B: WWA/B; strong wind A: SWA). The `weak wind' (WW) and `strong wind' (SW) runs had  $\chi=0.25,1$ and $\eta \equiv 2$ which resulted in an average speed of the wind of: 221.8 and 483.6 km/s, respectively.

Although the effects of galactic winds at $z>10$ is not clear
and very difficult to quantify we note that the feedback prescription
used here predicts a global star formation history and an IGM
(Intergalactic Medium) metal enrichment at $z=3$ that are in good
agreement with observations \citep{2002MNRAS.333..649S}.  The effect
of feedback by galactic winds on the IGM structures that surround
galaxies at $z\sim 3$ is extremely uncertain (e.g. \bcite{2003ApJ...584...45A}, \citeyear{2005ApJ...629..636A};
\bcite{2002ApJ...578L...5T}; \bcite{2004MNRAS.350..879D}; \bcite{2005ApJ...632...58R}), but we
stress that the speed of the wind values used in the feedback simulations
(although at much higher redshift) are in rough agreement with local
observations \citep{2005MNRAS.358.1453O}.

However, some caution is appropriate.
There can be many free unconstrained parameters and, with fine tuning,
one wonders if any set of observations could eventually be
reproduced. To attempt to assuage these fears and give more confidence
to the simulations, we have briefly summarised all of the free
parameters in the model and how they have been constrained in Table
\ref{tab:freeparameters}; the constraints are essentially the same as
those outlined in \citet{2002MNRAS.333..649S}. The simple star formation prescription (A) is not well motivated by observations, but serves to investigate the dependence of our results on the unknown star formation recipe. For prescription (B), there are a number of free parameters. However, notice that all of these are constrained by independent observations except for
the metallicity yield. We have not fine-tuned any of the parameters to achieve a particular result. This means that our model should produce believable physical results, with the exception of the absolute value (not the spread) of the metallicity. The absolute metallicity has been
`tuned' to fit the metallicity of the IGM at $z=3$ \citep{2002ApJ...578L...5T}. But for dwarf galaxies forming at $z=10$, it could conceivably be quite different. We comment further on this in section \ref{sec:metals}.

It is worth briefly noting some points of contention about the observations we use to constrain our model parameters. Firstly, we assume a Salpeter initial mass function (IMF) of stars. However, it is difficult to measure the massive end of the stellar mass function since these stars are so short-lived. Yet this is the region of most interest for determining the strength of the supernovae feedback and winds. Certainly the first-forming stars would appear, in theoretical models, to have an IMF biased towards the high-mass end \citep{1998ApJ...508..518A}. Secondly, our wind model posits that winds are driven during any period of star formation, even during relatively quiescent periods (recall that $\dot
M_{\rm w}=\eta \dot M_\star$). Alternatives link winds only to starbursts which are short-lived \citep{1999ApJ...513..142M}. There is, then, a potential danger in linking observations of star bursting dwarfs with our more continuous wind model. However, in the simulations, the majority of the stars form over a short timescale (see section \ref{sec:results}), so this should not be too great a concern. 

The galaxies were extracted from the simulation volume using a Friends-of-Friends algorithm with
linking length $l=0.2$. We follow the
iterative method of \cite{2002MNRAS.332..325P} to ensure that the
extracted halos are bound. At each stage the total energy of each
particle is calculated. Particles which do not appear to be bound are
excluded from the potential calculation in the next stage. The
momentum centre of the bound group is also calculated; this means
particles excluded at one stage may later re-enter the
calculation. We perform our analysis using only the bound component, although our tests
show the difference in profiles obtained is not significant.

\begin{table*}
\begin{center}
\setlength{\arrayrulewidth}{0.5mm}
\begin{tabular}{llll}
\hline
{\it S.F.P.} & {\it Free Parameter} & {\it Description} & {\it Constraint} \\
\hline
A & $\delta_c = 1000$ & Critical overdensity for star formation & Unconstrained \\
A & $T_c = 10^5$\,K & Critical temperature for star formation & Unconstrained \\
B & $\beta = 0.1$ & Mass fraction in stars $>8$M$_\odot$ & Salpeter IMF \\
B & $t_0^* = 2.1$\,Gyrs & Star formation time & Fit to \citet{1989ApJ...344..685K} law \\
B & $p = 3$Z$_\odot$ & Metallicity yield & Based on \citet{2002ApJ...578L...5T}; poorly constrained \\
\hline
A/B & $\epsilon_\mathrm{SN} = 10^{44}$\,J & Supernovae energy & Canonical value \\
A/B & $\eta = 2$ & Wind mass loss rate & Observed (\bcite{2005MNRAS.358.1453O}; \bcite{1999ApJ...513..156M}) \\
A/B & $v_\mathrm{wind} =$0, 221.8 and 483.6\,km/s & Wind speed & Observed (\bcite{2005MNRAS.358.1453O}; \bcite{1999ApJ...513..156M}) \\
\end{tabular}
\end{center}
\caption[]{Free parameters in the model and their observational constraints. Parameters are split into those which govern star formation (above) and those controlling the supernovae winds (below). The columns from left to right give the star formation prescription; free model parameters; a brief description of the free parameter; and observational constraints on that parameter. The metallicity is not constrained by independent observations (observations not of dwarf galaxies). We use a value based on \citet{2002ApJ...578L...5T}. They find $p=3$ provides the best fit to the metallicity of the IGM at $z=3$.}
\label{tab:freeparameters}
\end{table*}

\section{Results}\label{sec:results}

In this section we present the results from our suite of hydro-dynamical simulations. Recall that we have to stop these simulations at $z=10$ due to our small box size. This makes connecting the galaxies we form in our simulations to galaxies observed at $z=0$ difficult. The most massive galaxies we form have mass $\sim 10^8$M$_\odot$. From the linear theory arguments given in section \ref{sec:motivation}, we expect that $\sim 100$ of these galaxies will be accreted onto a galaxy the size of the Milky Way. Most will be accreted; ten percent will survive, as shown by \citet{2005astro.ph.10370M}. We can expect many of these survivors at $z=0$ to look very different from the galaxies we form at $z=10$. However, the smallest gas-poor galaxies in the LG at $z=0$ (the dSphs) are known to have very old stellar populations. As such, we can constrain our model by comparing the stellar distributions and metallicity of our most massive galaxies with data from the dSphs of the LG at the present epoch (see sections \ref{sec:stars} and \ref{sec:metals}). The connection to the dIrr galaxies will be more tentative, but we attempt this in section \ref{sec:mtol}.

\subsection{Overview}\label{sec:overview}

Figure \ref{fig:overview} shows the projected dark matter, star and gas densities for the second most massive halo in the NWA, NWB, WWA and WWB simulations. We do not show the results for the SW simulations since these are identical to the WW runs: the wind strength, for the plausible range of observed wind speeds, is not an important factor in our wind model. We present results for the SWA simulation alongside the others in section \ref{sec:mtol}. 

\begin{figure}
\begin{center}
\includegraphics[width=8.25cm]{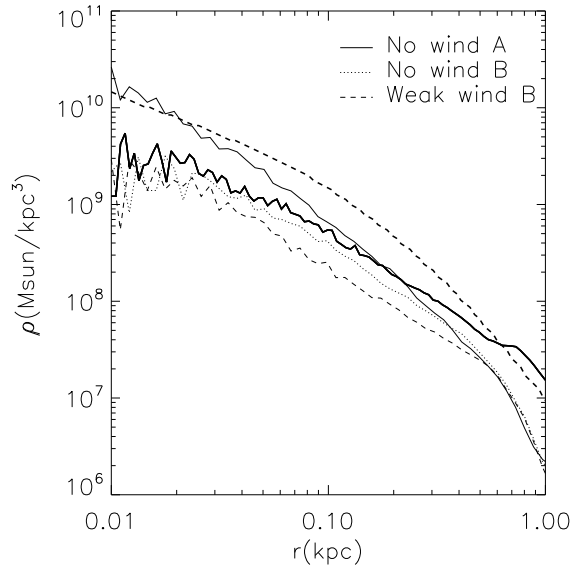}
\caption{Dark matter density profiles for the most massive halo in the NWA, NWB and WWB simulations. Over-plotted is the density profile of the most massive halo in the dark matter only simulation (thick black line), and a theoretical prediction for how this profile would respond to the adiabatic contraction of the baryons in the NWA simulation (thick dashed line).}
\label{fig:darkden}
\end{center}
\end{figure}

In the NWA simulation there is significant star formation; all of the most massive substructures and the central halo have bound stars. This recovers the familiar `over-cooling problem', which we discuss further in section \ref{sec:angmom}. There is very little gas left over and the gas density is two orders of magnitude lower than in all other cases. By contrast, the NWB, WWA and WWB simulations are similar: there is very little star formation and significant remaining gas. The remaining gas shows bubble-like regions with over-densities of a factor of $\sim 20$. These over-densities are caused by heating from the star forming regions. In the case of the WWA model, this is a result of the supernovae driven galactic winds; for the NWB model, it is the result of heating from the supernovae feedback. 

\begin{figure*}
\begin{center}
\includegraphics[width=5.8cm]{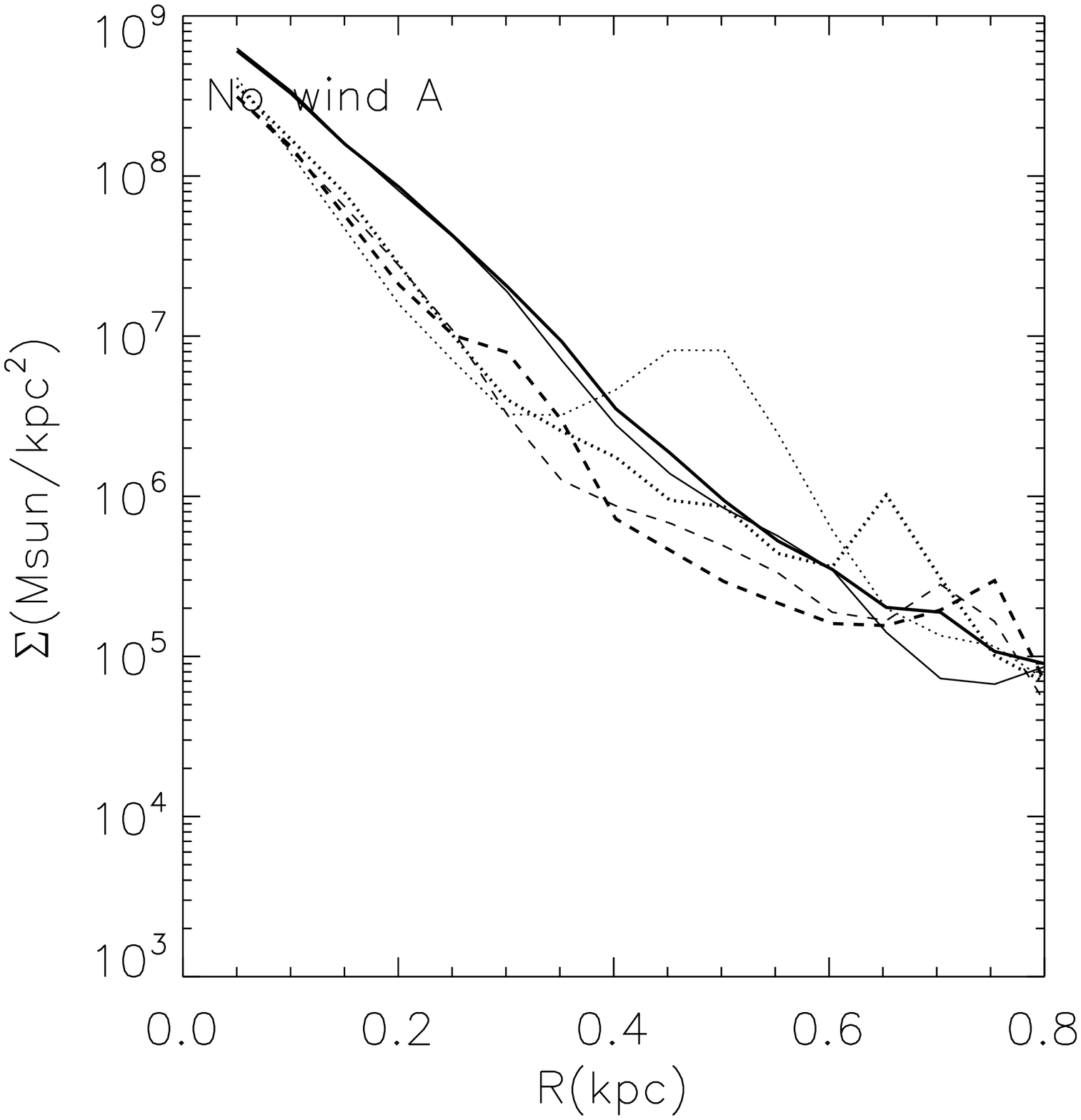}
\includegraphics[width=5.8cm]{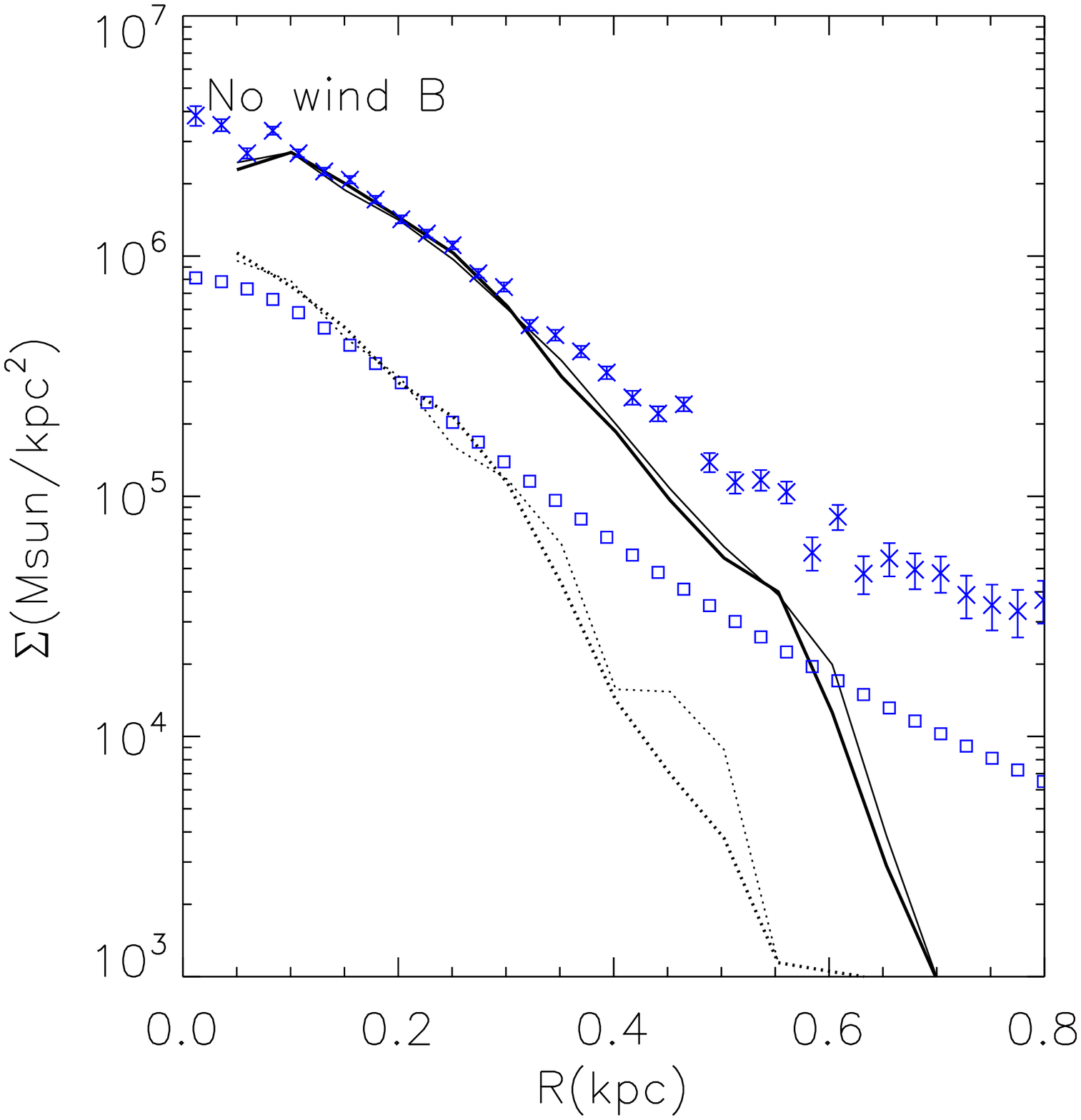}
\includegraphics[width=5.8cm]{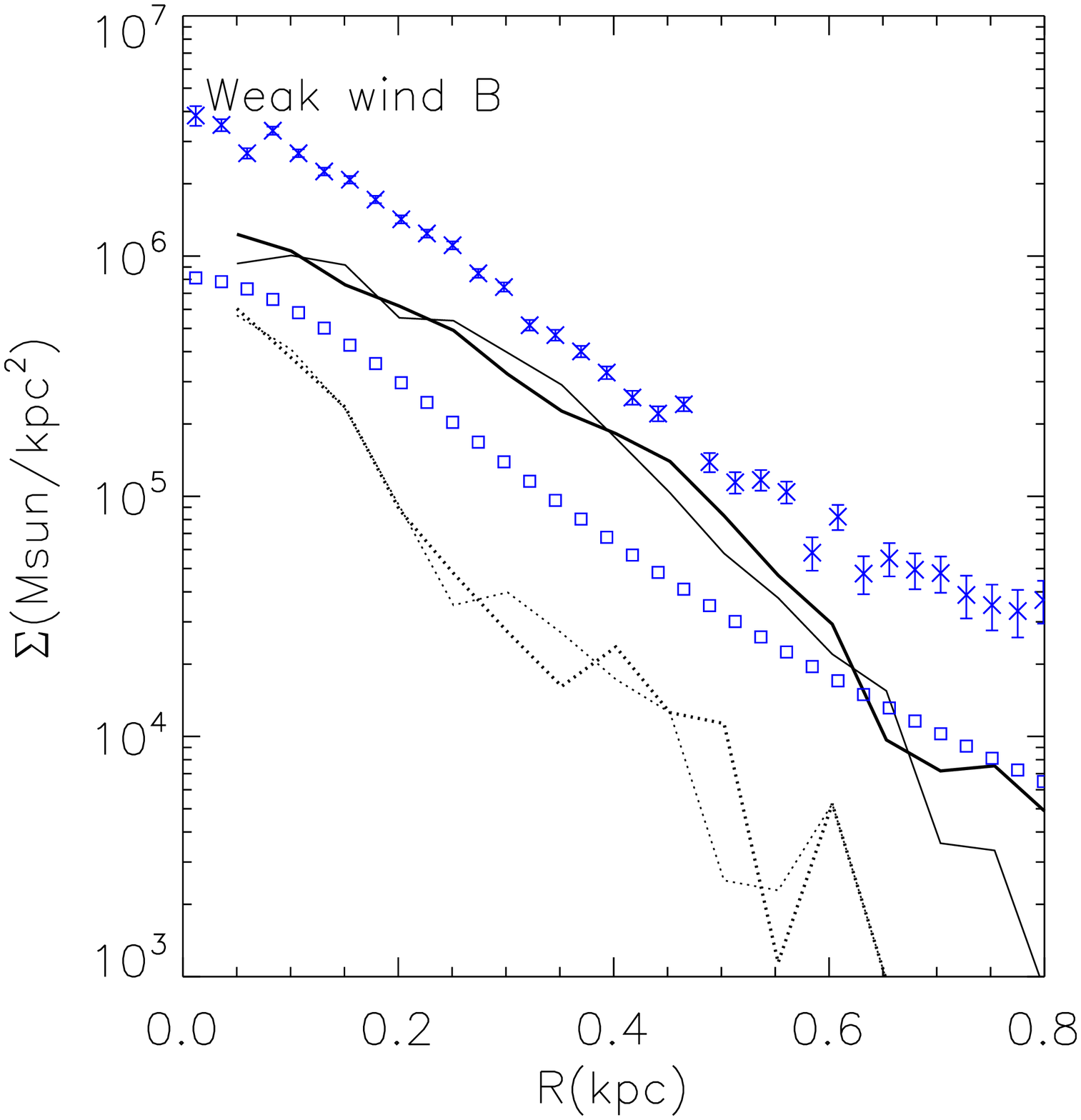}\\
\includegraphics[width=5.8cm]{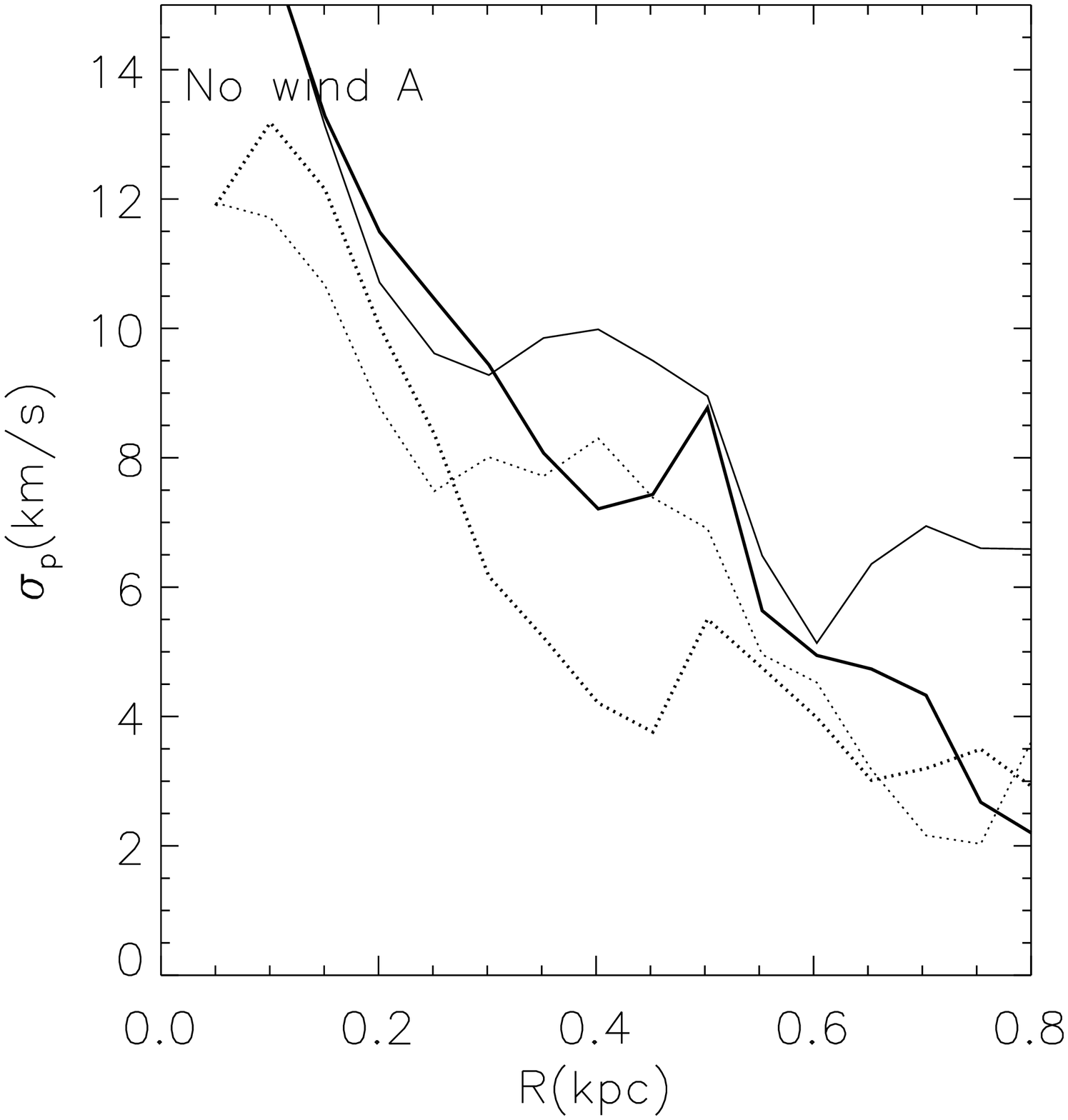}
\includegraphics[width=5.8cm]{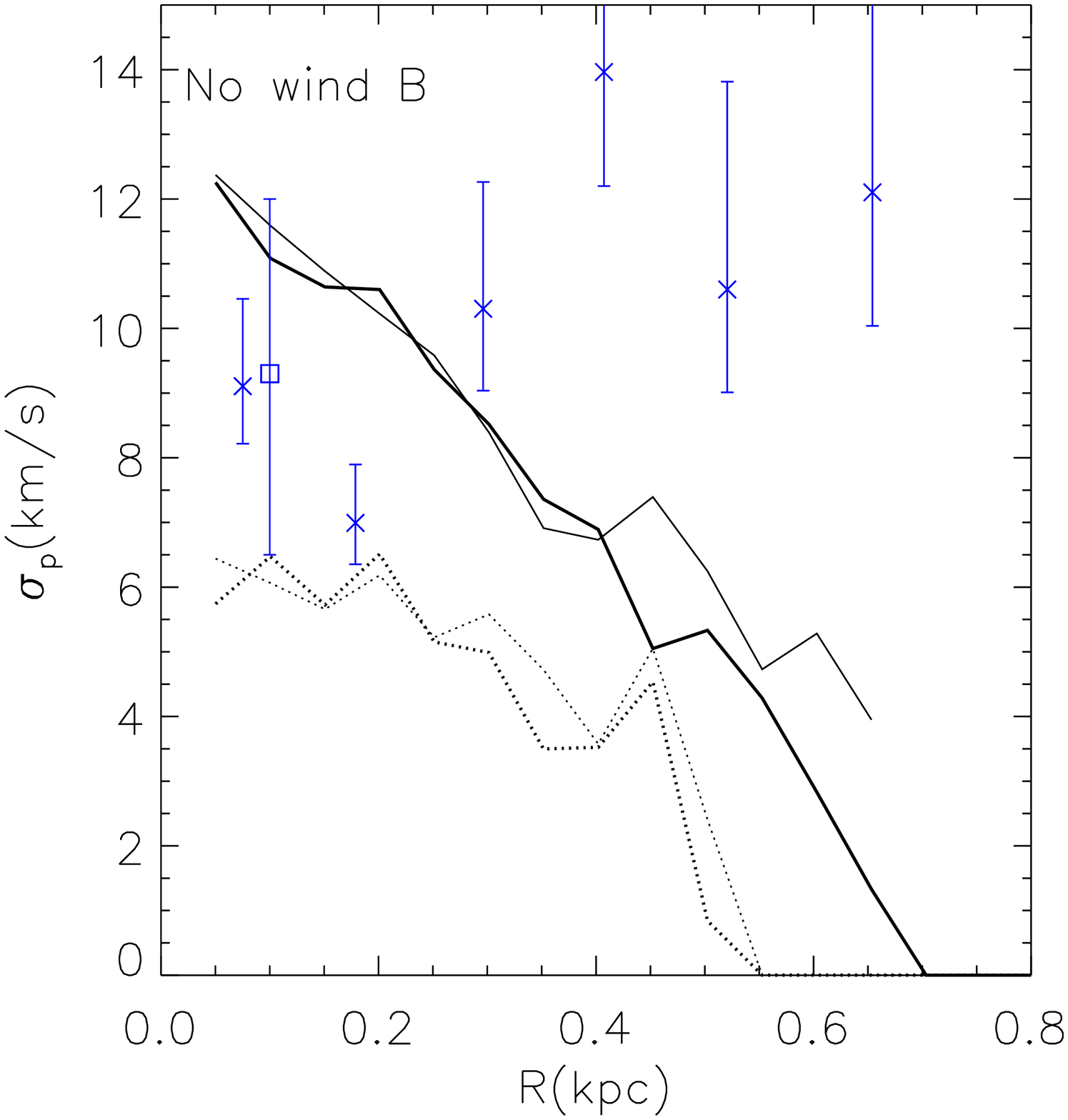}
\includegraphics[width=5.8cm]{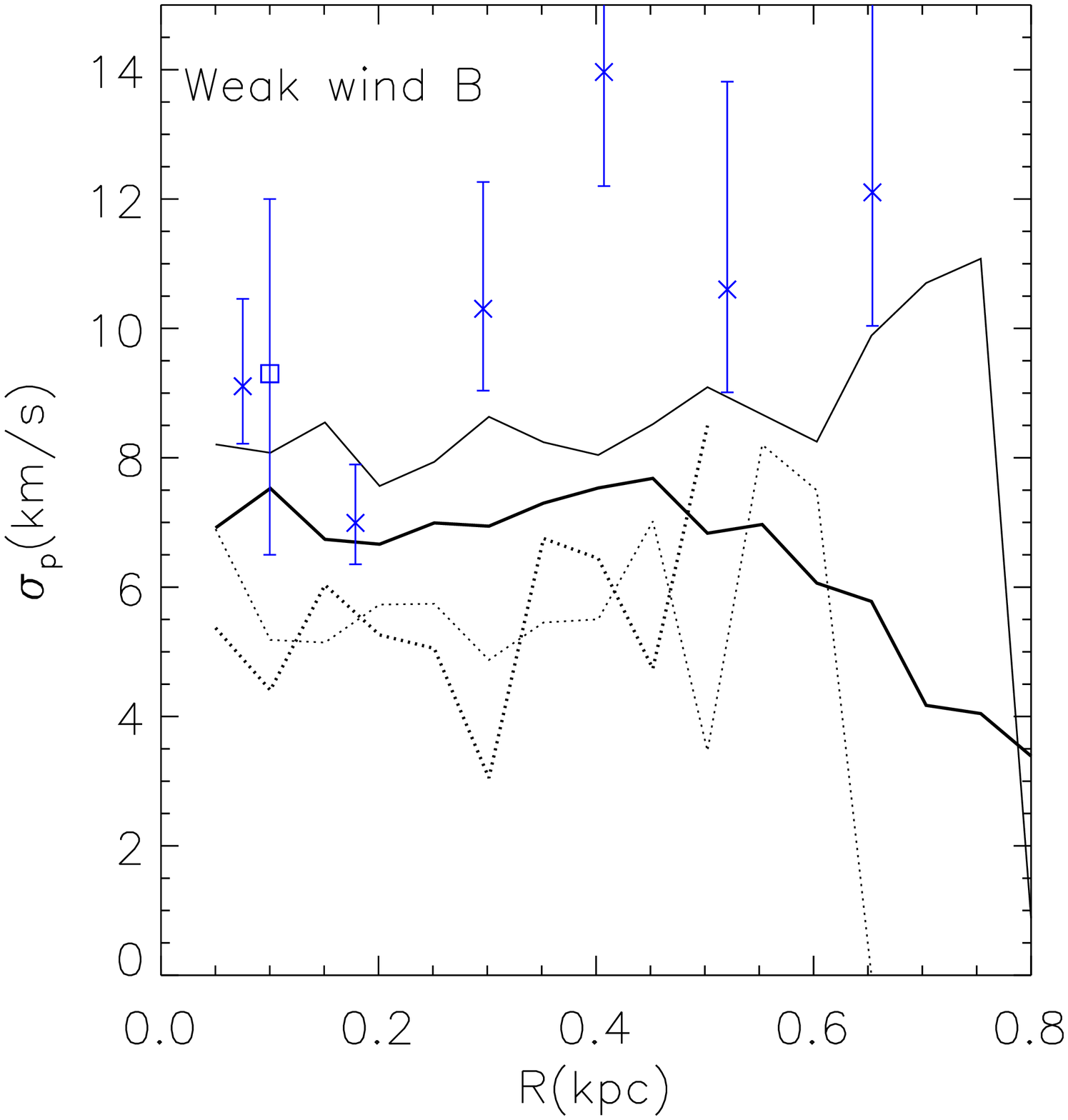}
\caption{Projected surface density (top) and velocity dispersion (bottom) for stars in the two most massive halos (solid and dotted lines) in the NWA, NWB and WWB simulations. The thick and thin lines show two different projection angles which bracket the range of projected profiles. Over-plotted are data from two LG dSphs which bracket the observed range of brightness: Draco (crosses) and, the recently discovered, Ursa Major (squares). The data were taken from \citet{2001ApJ...563L.115K}, \citet{2004ApJ...611L..21W}, \citet{2005ApJ...626L..85W} and \citet{2005ApJ...630L.141K}. Ursa Major is so faint that it has only one data point for the kinematics, but this suggests a very similar central velocity dispersion to Draco, despite it being an order of magnitude fainter. For Ursa Major's surface brightness distribution, we use the measured value for the scale length and central brightness, but assume the same distribution as in Draco - hence the lack of error bars for these points. Note that the data is from dSph galaxies at redshift $z=0$, while the simulations stop at $z=10$. The comparison is only meaningful because the dSph stellar populations are old (see start of this section for further details).}
\label{fig:stars}
\end{center}
\end{figure*}

The key point to take away from Figure \ref{fig:overview} is that all of the simulations with feedback produce qualitatively similar results: suppressed and spatially extended star formation. In particular, the NWB and WWA simulations are almost identical in every respect despite their quite different star formation and feedback mechanisms. For this reason, we plot only the results for the NWB simulation from here on. We come to why the NWB and WWA simulations are so similar in section \ref{sec:angmom}. None of the simulations result in a significant `blow out' of gas, even in the galactic wind models. This is to be expected from the analytic arguments given in section \ref{sec:motivation}. 

Recall that we used identical initial phase space conditions for each of the simulations. We plot identical halos which differ only in their star formation and feedback prescriptions. Yet the evolution of substructure in each of the simulations is quite different. The dark matter halo in the NWA simulation is much more centrally concentrated. This is a result of the increased merging due to the condensation of a large mass fraction of baryons at the centre of each substructure halo. By contrast, the WWA and NWB simulations are much less concentrated, corresponding to less merging. Finally, the WWB simulation (which has the strongest feedback - a combination of supernovae heating and galactic winds) shows the least merging of all. Substructure can still clearly be seen to the bottom left of the main halo, both in the dark matter and gas distributions, and below right in the dark matter. 

The dissipation and collapse of gas significantly alters the merger history of halos (c.f. \bcite{2006MNRAS.366.1529M}). However, it does not alter the dark matter density profiles, just the concentration. In Figure \ref{fig:darkden}, we plot the dark matter density profiles for the most massive halo in the NWA, NWB and WWB simulations. Over-plotted is the density profile of the most massive halo in the dark matter only simulation (thick black line), and a theoretical prediction for how this profile would respond to the adiabatic contraction of the baryons in the NWA simulation (thick dashed line). For this analytic calculation, we used the usual prescription given in \citet{1986ApJ...301...27B}. This assumes that the halo collapses spherically without shell crossing, conserving mass and specific angular momentum and with all halo particles moving on circular orbits. It provides a reasonable fit in the central regions to the NWA simulation, but performs poorly over intermediate radii. This is a known result. Even before the Blumenthal paper, \citet{1980ApJ...242.1232Y} demonstrated analytically that the assumption that the halo particles move on circular orbits leads to an over-estimation of the contraction. More recently, \citet{2004ApJ...616...16G} have used simulations to demonstrate the same disparity on galaxy cluster scales. Here we verify that the problem persists on the scale of dwarf galaxies. The standard adiabatic contraction model should be used with caution on these scales. With supernovae feedback there is a {\it lower} central halo density than in the dark matter only run, while without feedback, the Blumenthal model over-predicts the effect. This agrees well with previous findings in the literature (\bcite{2002MNRAS.333..299G}; \bcite{2005MNRAS.356..107R}). All of the dark matter halos are well-fit by NFW profiles \citep{1996ApJ...462..563N}; the typical concentration parameter and mass are $c \sim 4$, $M \sim 10^8$\,M$_\odot$. Even in the presence of a strong wind, a central dark matter cusp persists. 

\begin{figure*}
\begin{center}
\includegraphics[width=5.8cm]{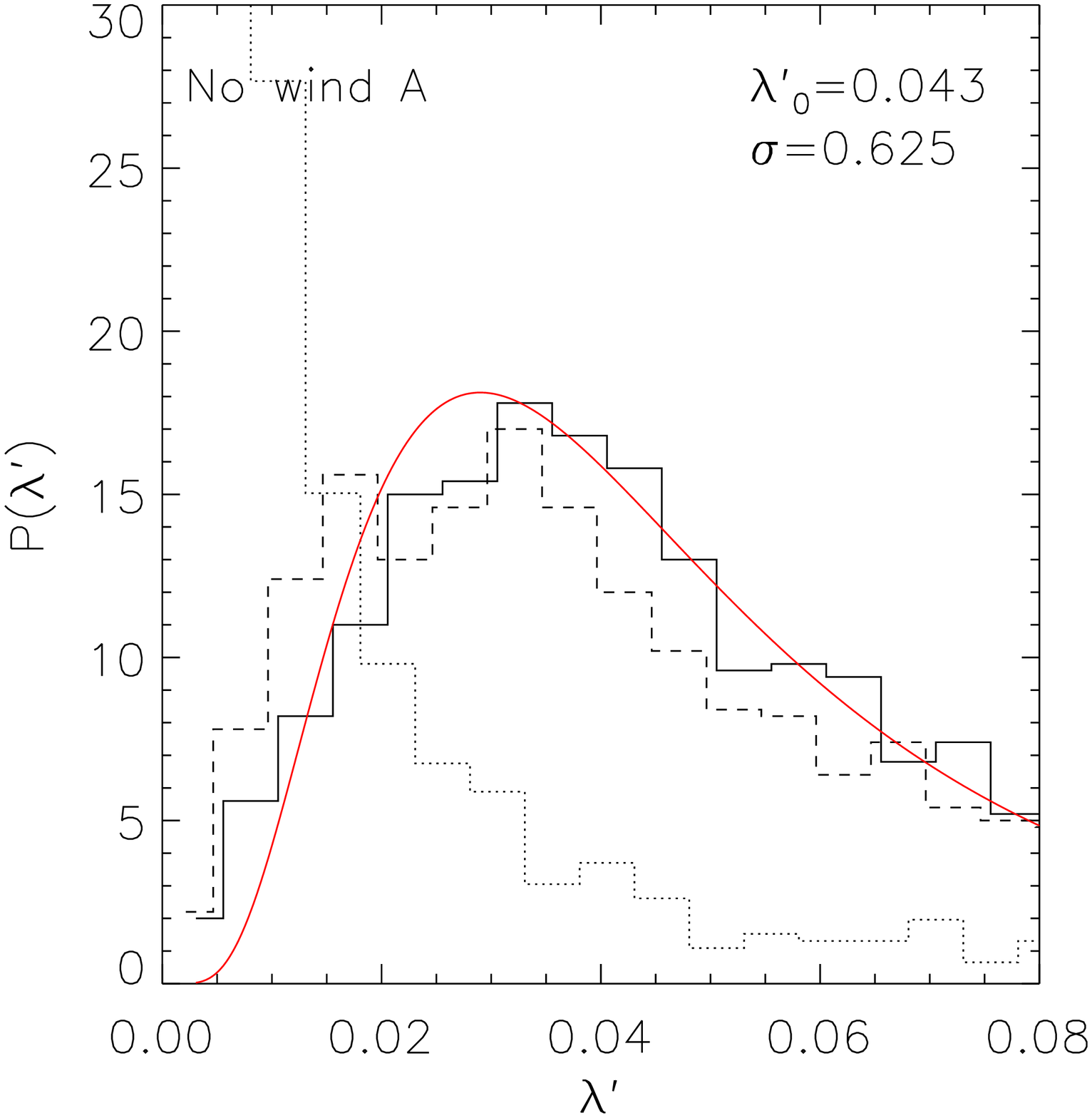}
\includegraphics[width=5.8cm]{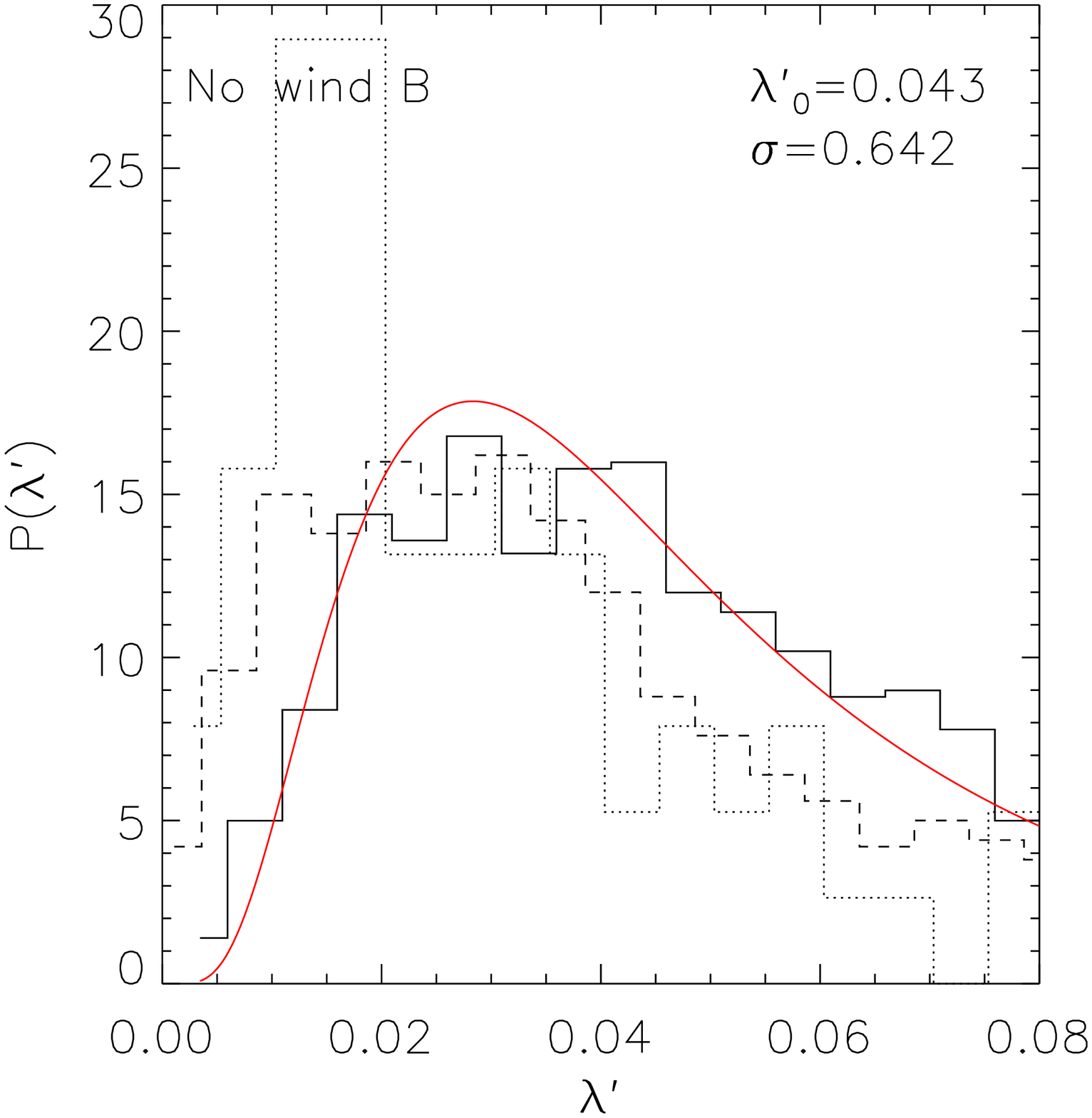}
\includegraphics[width=5.8cm]{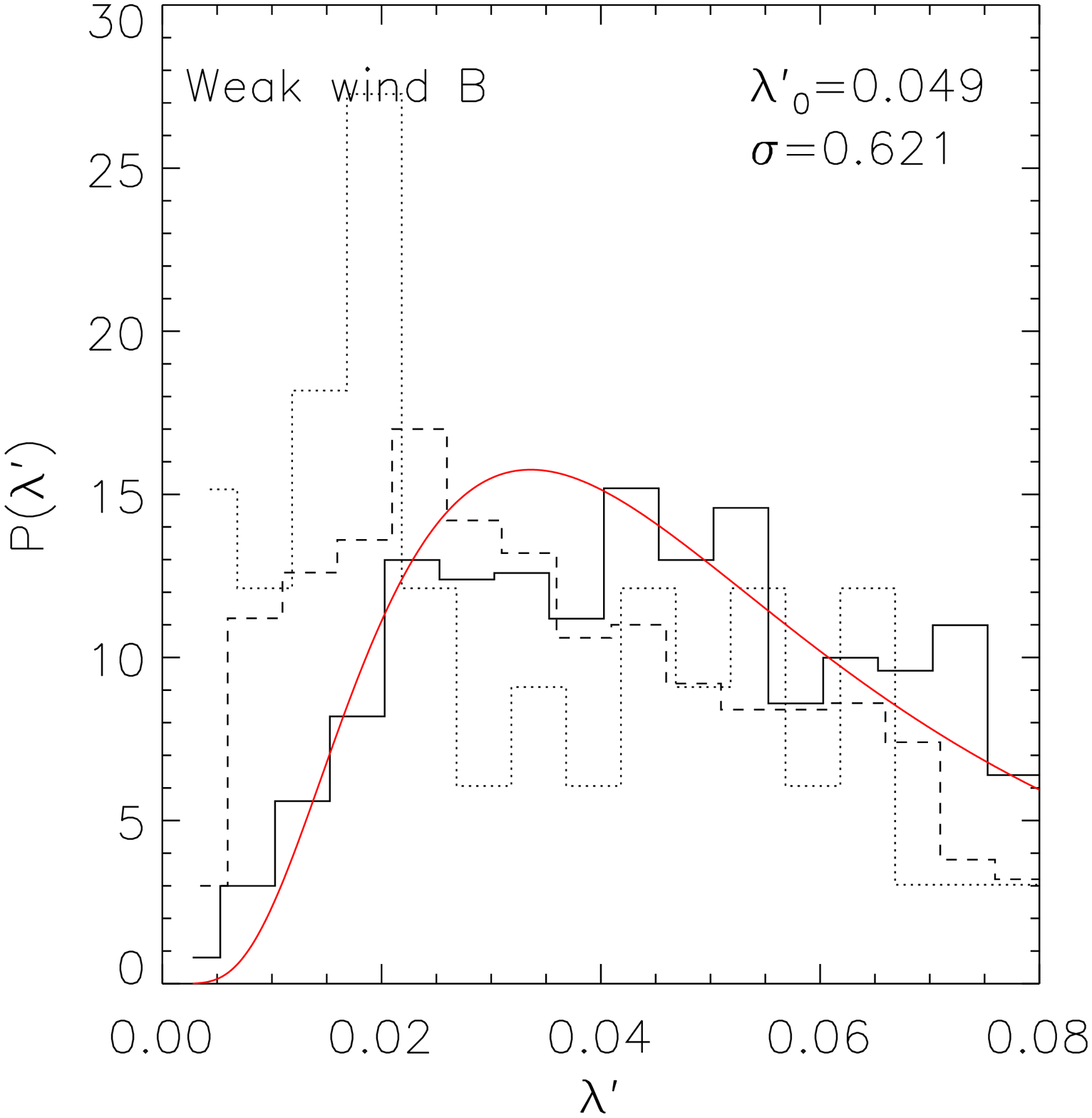}
\caption{Distribution of spin parameters ($\lambda'$) for the NWA, NWB and WWB simulations. The solid, dotted and dashed histograms show $\lambda'$ for the dark matter, stars and gas. The smooth solid line is a lognormal fit to the dark matter $\lambda'$ (see equation \ref{eqn:lognormal}). The best-fit parameters are marked on each plot.}
\label{fig:angmom}
\end{center}
\end{figure*}

\subsection{Stars viewed in projection}\label{sec:stars}

Figure \ref{fig:stars} shows the projected surface density (top) and velocity dispersion (bottom) for the stars in the two most massive halos (solid, dotted and dashed lines) in the NWA, NWB and WWB simulations. The thick and thin lines show two different projection angles. Over-plotted are data from two LG dSphs which bracket the observed range of brightness: Draco (crosses) and, the recently discovered, Ursa Major (squares). The data were taken from \citet{2001ApJ...563L.115K}, \citet{2004ApJ...611L..21W}, \citet{2005ApJ...626L..85W} and \citet{2005ApJ...630L.141K}. Ursa Major is so faint that it has only one data point for the kinematics, but this suggests a very similar central velocity dispersion to Draco, despite it being an order of magnitude fainter. For Ursa Major's surface brightness distribution, we use the measured value for the scale length and central brightness, but assume the same distribution as in Draco - hence the lack of error bars for these points. 

In the NWA simulation a significant amount of substructure within the main halo forms stars (see Figure \ref{fig:overview}). This can clearly be seen in the surface density and projected velocity dispersion of the stars, both of which, for the second most massive halo, show a significant bump correlated with the position of the largest substructure halo. By contrast the NWB simulation, in which the substructure halos form few or no stars, show no such features. Such substructure is interesting. Several of the LG dwarfs appear to have bumps and wiggles in their projected velocity dispersion profiles. For two, the projected velocity dispersion appears to fall sharply towards the edge of the light, which is dropping off smoothly \citep{2004ApJ...611L..21W}. For the WWA simulation, we did find one halo which retained a few stars within a subhalo. This can be seen in Figure \ref{fig:overview}: notice for the stars in the WWA simulation, there is an overdensity to the bottom left correlated with a dark matter subhalo. These subhalo stars also lead to a sharp drop in the projected velocity dispersion along some lines of sight, with no associated sharp truncation in the stellar surface density. 

Such substructures are unlikely to be very long-lived. However, it is interesting to speculate that  kinematic bumps and wiggles could correspond to {\it late-infalling} substructure within the dwarfs. 

The NWA simulation halos have very high stellar surface densities and consistently falling projected velocity dispersions, whereas the LG dSphs have low surface densities and flat projected velocity dispersions (see data points, middle panels). This highlights the importance of supernovae heating and galactic winds for keeping the surface density of the stars low and producing galaxies which qualitatively resemble the dSphs of the LG.  

The NWB halo much more closely resembles the LG dwarfs in their stellar distribution (see middle panels). The agreement inside $\sim 400$\,pc is very encouraging, especially given that we have made no attempt to fine tune our model to fit the data. However, for our simulated galaxies, the projected surface brightness and velocity dispersion both fall too steeply beyond $\sim 400$\,pc. 

The WWB model - which includes both supernovae heating and galactic winds - seems to solve these problems. The stars are more extended and the velocity dispersions are flat. This is exactly what we expect from gas mass loss. For the NWB  simulation, the gas contributes significantly to the potential from $\sim 0.4$\,kpc outwards. This explains why the stellar velocity dispersion is falling rather than flat (as would be expected if only the dark matter halo contributed to the potential). As shown in \citet{2005MNRAS.356..107R} (and see also \bcite{2005ApJ...624..726M}) gas mass loss will cause the remaining stars to expand and settle into an approximately exponential distribution - similar to that observed in Draco. Since the final potential is then dominated by the dark matter, the resulting velocity dispersions are flat. 

However, it is important to remember that we are comparing galaxies forming at $z=10$ with those observed at $z=0$ in the Local Group. As such, the above discrepancies could also be the natural result of {\it external} feedback which is not included in our model. The tidal field from a large nearby galaxy, like the MW, will heat stars beyond a characteristic `tidal radius' ($r_t$), leading to more extended surface brightness profiles and flat or rising projected velocity dispersions beyond $r_t$ \citep{2006MNRAS.tmp..153R}. Using the analytic formulae for $r_t$ given in \citet{2006MNRAS.366..429R}, we find for a Draco of total mass $10^8$M$_\odot$, with a tidal radius of $r_t = 400$\,pc, that its orbital pericentre is $\sim 20$\,kpc. This is consistent with measurements of the proper motion of Draco \citep{2001ApJ...563L.115K}. An alternative possibility is gas mass loss due to ram pressure stripping, which is likely to mimic the effect of the galactic wind in WWB. We return to these issues in section \ref{sec:mtol}.

\begin{figure}
\begin{center}
\includegraphics[width=8.25cm]{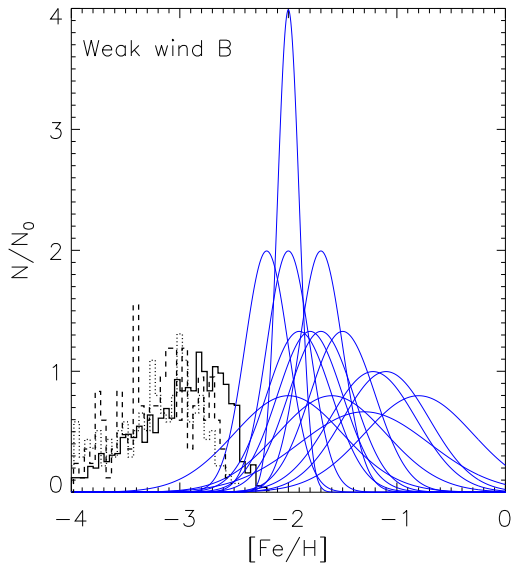}
\caption{Metallicity distributions of the three most massive halos in the WWB simulation (thick solid, dotted and dashed histograms). The smooth solid lines show data for the LG dSphs taken from \citet{1998ARA&A..36..435M}. We have assumed that these distributions are Gaussian, although detailed recent observations suggest more complex distributions are likely (see e.g. \bcite{2004ApJ...617L.119T} and \bcite{2006AJ....131..895K}).}
\label{fig:metals}
\end{center}
\end{figure}

\begin{figure}
\begin{center}
\includegraphics[width=8.25cm]{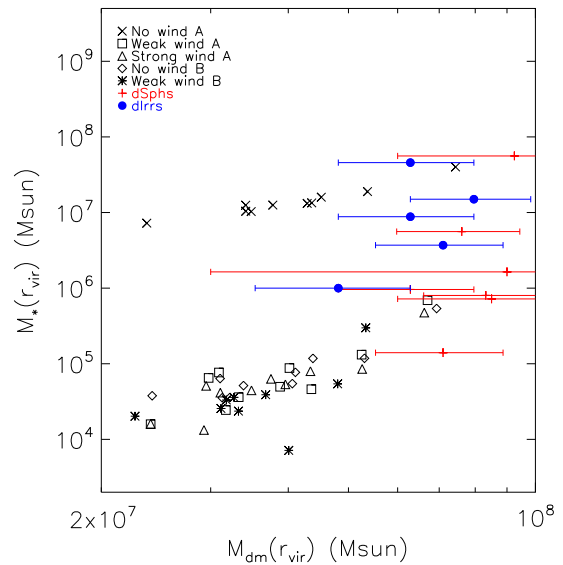}
\caption{Total stellar mass as a function of dark matter mass within the virial radius for the ten most massive halos in the NWA, NWB, WWA, WWB and SWA simulations. Over-plotted are data for the LG dwarfs and nearby dIrrs. See text for details of the data compilation.}
\label{fig:mtol}
\end{center}
\end{figure}

\subsection{Angular momentum and star formation}\label{sec:angmom}

The angular momentum of a halo is typically parameterised by the dimensionless spin parameter given by \citep{2001ApJ...555..240B}:

\begin{equation}
\lambda'_i = \frac{J_i}{\sqrt{2}M_i V_\mathrm{vir}R_\mathrm{vir}}
\label{eqn:lambdadash}
\end{equation}
\begin{equation}
J_i  = |\sum_{j=0}^N m_{i,j} \hspace{1mm} \underline{r}_j \times \underline{v}_j|
\label{eqn:Ji}
\end{equation}
where the subscript, $i$, denotes the particle species (dark matter, stars or gas), $J_i$ is the total angular momentum in that species, $R_\mathrm{vir}$ is the virial radius\footnote{The virial radius is here defined as the radius within which the mean density of the halo is equal to 200 times the critical density of the universe at $z=10$.}, $M_i$ is the mass of that species interior to $R_\mathrm{vir}$ and $V_\mathrm{vir} = \sqrt{GM_\mathrm{vir}/R_\mathrm{vir}}$ is the circular speed at the virial radius. This definition means that $\lambda'_i = 1/\sqrt{2}$ if all of the particles are orbiting on circular orbits at $R_\mathrm{vir}$.

\citet{2001ApJ...555..240B} demonstrated that the lognormal profile provides an excellent fit to the distribution of spin parameters in their simulations. \citet{Colin:2003jd} showed that the same form provides an excellent fit to dwarf galaxy scale halos. The lognormal profile is given by:

\begin{equation}
P(\lambda') = \frac{1}{\lambda'\sqrt{2\pi \sigma}}\exp\left(-\frac{\ln^2(\lambda'/\lambda'_0)}{2\sigma^2}\right)
\label{eqn:lognormal}
\end{equation}

In Figure \ref{fig:angmom}, we plot the distribution of spin parameters ($\lambda'$) for the NWA, NWB and WWB simulations. The solid, dotted and dashed histograms are for the dark matter, stars and gas. The smooth solid line shows the log normal fit to the dark matter; the best fit parameters are shown in the top-right of each plot. These are in excellent agreement with those found by \citet{Colin:2003jd}. In each case we used all halos with more than $\sim 10$ particles (in dark matter, stars and gas respectively), which is about $\sim 1000$ halos per simulation. 

For the NWA simulation, we recover the well-known over-cooling problem. The simple star formation prescription we use turns all of the cold gas into stars and over-merging leads to a very low final stellar angular momentum. The NWB simulation alleviates this problem, but low angular momentum stars still remain. The WWB simulation, with the strongest feedback, prevents almost any angular momentum loss from the stars. As expected analytically (see e.g. \bcite{2002MNRAS.335..487M}), feedback solves the over-cooling problem by preventing the subhalos from efficiently forming stars. However, we should be a little cautious. \citet{2006astro.ph..1115K} have recently shown that, for {\it disc galaxies}, $> 10^6$ particles are required {\it per galaxy} for angular momentum loss due to spurious numerical transport to be at an acceptable level. Even with our very high resolution, our best resolved halos have O$(10^5)$ particles -- an order of magnitude too low to avoid such numerical over-cooling. The angular momentum loss mechanisms identified by \citet{2006astro.ph..1115K} are: viscous friction between the cold disc and hot halo; transfer between the spiral arms and bar of the disc and the dark matter halo; and angular momentum loss from infalling cold gas clouds. In our simulations, none of these mechanisms can act. In the simulation without supernovae feedback (NWA), all subhalos turn their gas into stars. In this case, discs do not form because of angular momentum loss in mergers. With supernovae feedback (NWB, WWA, WWB, SWA), we prevent stars from forming in the smaller subhalos. Now discs do not form because the velocity dispersion in the gas, maintained by the feedback at $\sim 10$\,km/s, is comparable to the rotational velocity. It is essentially a disc with a very large scale height. We can see this in Figure \ref{fig:angmom}, where in all of the simulations with supernovae feedback the stars and gas retain their initial angular momentum, but do not significantly condense. For these simulations, with supernovae feedback, stars form stochastically where the density fluctuates upwards. It is likely that such fluctuations are imprinted by numerical noise, but they can be thought of as physical. In the `real' universe, such fluctuations are likely to exist as a result of external tides and internal micro-physics.

The above explains why the feedback recipe is not critical in determining the final morphology of the galaxy. Provided that it prevents the subhalos from efficiently forming stars then it will work. Galactic winds do this by ejecting a proportion of the gas available for star formation; feedback from supernovae heating does this by heating locally cold gas which would otherwise form stars. This is a known result. For example, in \citet{2000MNRAS.315L..18E}, they show that the over-cooling problem can be solved by simply delaying star formation until after most subhalos have merged into the main galaxy. 

Finally, note that while in the NWB and WWB simulations the stars retain significant angular momentum with $\lambda'_0 \sim 0.05$, this would be very hard to detect in practise. None of the galaxies show measurable tip-to-tip rotation of greater than $\sim 2$km/s. This is because the stars remain extended and do not collapse and `spin up'.

\subsection{Metallicities}\label{sec:metals}

In Figure \ref{fig:metals} we plot the metallicity distributions of the three most massive halos in the WWB simulation (solid, dotted and dashed histograms); the NWB simulation showed very similar results (recall that the simple star formation prescription - A - does not produce reliable metallicity output). The smooth solid lines show data for the LG dSphs taken from \citet{1998ARA&A..36..435M}. We have assumed that these distributions are Gaussian, although detailed recent observations suggest more complex distributions are likely (see e.g. \bcite{2004ApJ...617L.119T} and \bcite{2006AJ....131..895K}).

Recall from section \ref{sec:hydro} that the absolute value of the
metallicity for our halos should be taken with caution as we have chosen a yield which produces a good fit to the metallicity of the IGM at $z=3$ \citep{2002ApJ...578L...5T}. It is not clear that this is the relevant value for our dwarf galaxies forming at $z=10$. That caveat aside, our metallicities are systematically lower than the observed dwarfs. This is, perhaps, to be expected: the galaxies we form in WWB seem to be less massive, both in stars and dynamically, than most observed nearby dwarfs (see Figure \ref{fig:stars} and Figure \ref{fig:mtol} in the following section). Recent observations of the {\it faintest} and lowest mass dwarf galaxies suggest metallicities which are lower than the mean of those shown in Figure \ref{fig:metals}. The faintest galaxy to date - Bootes - has a mean metallicity of $\sim -2.5$ \citep{Munoz:2006vg}, closer to the most massive galaxy we form in WWB (especially considering that the yield is so poorly constrained). 

The spread, which is more certain in our models, appears to match quite well that of the nearby dwarfs. But in our models it is clearly non-Gaussian (recall that we assumed Gaussianity for the data). The shape of our distribution matches very well that of more detailed modelling recently performed by \citet{Marcolini:2006mk}. We find, similar to their results, an asymmetric distribution peaked at high metallicity, with a tail to lower metallicities. Recent work by \citet{2002AJ....124.3222B} suggests such asymmetries may be present in the Draco and UMi dSphs, but it would be interesting to confirm this as a generic feature in  future observational work. 

Our model may provide a simple solution to the `abundance problem' for the LG dwarfs. \citet{2003AJ....125..707T} found, from detailed spectroscopy of resolved stars, that: 1) the abundance\footnotemark of stars in four LG dSphs is lower than that observed in the old stars of the MW stellar halo; and 2) the mean metallicity of the MW halo stars is much lower than that of the dwarfs: $<$[Fe/H]$>\sim -3$. This presents a puzzle since it suggests that the MW stellar halo cannot have formed from galaxies like the LG dSphs seen at the present epoch. In our model this is not a problem. Our progenitor `baryonic building blocks' (BBBs) do have the correct mean metallicity. Furthermore, while we do not explicitly track abundances in our code, we can expect their abundances to be high. This is because there is a link between abundance and the star formation timescale. Stars which form rapidly are only enriched by Type II supernovae which do not produce much iron; stars which form more slowly can be enriched by Type Ia supernovae which produce more iron relative to other metals and therefore lower the abundance (see e.g. \bcite{1998gaas.book.....B}). Our early forming BBBs form stars rapidly and are likely to be of high abundance. Over a Hubble time, as our BBBs gradually form more stars, they then lower their abundance and start to look more like the LG dSphs. This agrees well with earlier studies in the literature (\bcite{2005ApJ...635..931B}, \bcite{2005ApJ...632..872R} and \bcite{2006ApJ...638..585F}). 

\footnotetext{The abundance is the mass of a given `metal' (element heavier than Helium) relative to iron.}

\subsection{Mass to light ratios}\label{sec:mtol}

In Figure \ref{fig:mtol}, we plot the total stellar mass as a function of dark matter mass within the virial radius for the ten most massive halos in the NWA, NWB, WWA, WWB and SWA simulations. We stop at the first ten halos since the smallest of these halos have only $10^4$\,M$_\odot$ in stars which corresponds to just $\sim 10$ star particles. Over-plotted are data for the LG dwarfs and nearby dIrrs. 

It is difficult to obtain an accurate compilation for nearby dwarf galaxies. The dIrrs have only gas kinematic measurements from which the dynamical mass can be derived; while the dSphs have only stellar kinematics (since they are devoid of gas). A further complication comes from the quality of the older versus the newer data. Some of the dSphs now have excellent data (like that over-plotted in Figure \ref{fig:stars}). But most have older data which was typically taken only near the centre of the galaxy. It is now known that these older measurements systematically underestimate the mass of the dSph galaxies. Our adopted solution is to use only the latest data. These have been compiled from \citet{2004A&A...413..525B}, \citet{2003A&A...409..879B}, \citet{2006MNRAS.365.1220B}, \citet{2005ApJ...630L.141K}, \citet{2005ApJ...626L..85W}, \citet{2005ApJ...631L.137M}, \citet{2004MNRAS.354L..66K}, \citet{2004ApJ...611L..21W}, \citet{2001ApJ...563L.115K}, \citet{2005astro.ph.11465W}, \citet{2006astro.ph..3694W}, \citet{1995MNRAS.277.1354I} and \citet{2006AJ....131..375W}. For the dSphs, we plot the masses derived from detailed mass modelling in all cases except for Ursa Major (the faintest dSph)\footnotemark. For this dSph and for the dIrrs we assume an isotropic isothermal sphere model. This gives:

\begin{equation}
M(r) = \frac{3 f \overline{\sigma_p^2} r}{G}
\label{eqn:isothermal}
\end{equation}
where $G$ is the gravitational constant, $\overline{\sigma_p^2}$ the measured {\it mean projected} velocity dispersion squared, $r = 1$\,kpc $\sim R_\mathrm{vir}$ is the radius enclosing the mass and $f$ is a small correction factor. $f = 1.4$ is chosen such that we would recover the correct mass for the Draco dSph (obtained from detailed distribution function modelling) using equation \ref{eqn:isothermal}.

\footnotetext{Ursa Major has only one velocity dispersion measurement. It could be reasonably argued that this makes it as unreliable as the older data. However, unlike the older data, this one data point lies $\sim 250$\,pc away from the centre of the galaxy. This makes it more likely to be representative of the mean velocity dispersion than older measurements.}

There are two key things to take away from Figure \ref{fig:mtol}. Firstly, notice how
rapidly the stellar mass falls as the halo mass is reduced in all of the simulations with internal feedback from supernovae. The details of the star formation and feedback recipes are not important for determining the final stellar masses. In all cases, the stellar mass falls two orders of magnitude for
a decrease in halo mass of little over 30\%. This was expected from the analytic arguments presented in section \ref{sec:motivation}. Secondly, notice that the LG dwarfs and nearby dIrrs seem to show a cut-off in their dynamical mass at $\sim 5 \times 10^7$M$_\odot$, but a wide range in stellar masses. The data are well bracketed by our NWA and other simulations. 

The dSph galaxies are consistent with our most massive halos with all of the gas removed; while the dIrr lie close to the NWA model suggesting that, over a Hubble time, they have managed to turn most of their initial gas into stars. That this process is a slow one is important. Firstly, because star formation is inefficient, it allows gas to be slowly removed. Ram pressure stripping is one such mechanism which could achieve this (see e.g. \bcite{2005astro.ph..4277M}). It has the advantage that galaxies which lie close to the MW or M31 will be more rapidly stripped than those which lie further away, naturally reproducing the distance morphology relation (see section \ref{sec:introduction}). Secondly, recall that in the NWA simulation, the stellar density was much higher than that observed in the LG dwarfs and too much substructure formed stars. This occurred because the star formation in the NWA simulation was too efficient. Even for the most isolated dIrr which eventually turns all of its gas into stars, supernova feedback and reionisation have a role to play in keeping the star formation efficiency low and spatially extended. This prevents substructure forming stars and keeps the stellar surface density and rotational velocity low, consistent with observations.

One caveat is worth mentioning. In all cases, it is our most massive simulated galaxies which resemble the {\it least massive} LG dSphs. For the WWB simulation, none of our simulated galaxies achieve enough stellar mass to be consistent with observations (this can also be seen in Figure \ref{fig:stars}). This suggests that either our feedback prescription is too strong, or the observed dwarfs formed from the merger of several of our BBBs, with some continued star formation over a Hubble time. It is not possible to address which of these is correct without running larger-box simulations of comparable resolution and continuing down to lower redshift. We leave this for future work. 

\section{Discussion}\label{sec:discussion}

\subsection{A connection to galaxies in clusters?}

\citet{2002MNRAS.333..423T} have recently discovered a new class of very low surface brightness (VLSB)  galaxies in the Virgo cluster. These galaxies are extremely faint, spheroidal in morphology and extended over several kpc. So far in this paper, we have talked only about the low density environment of the LG. In clusters, the picture could be quite different. While little is known about these VLSB galaxies and follow-up observations are currently underway, an intriguing possibility is that these too are naked stellar halos. Like the dSph galaxies of the LG, they may have lost their gas faster than they could turn it into stars, leaving them with just the extended, low rotational velocity, metal poor, old stellar halo component. If true, further observations should confirm that these VLSB galaxies have all of the properties of old stellar halos. They are likely more massive than their LG cousins, however. 

\subsection{The formation of globular clusters}

We have presented the case for a `baryonic building block' for galaxy formation. These gas-rich building blocks have total mass $M_\mathrm{crit} \sim 10^8$M$_\odot$ and form $\sim 10^6$M$_\odot$ in stars before reionisation. However, we can see in the universe that stars form on smaller scales than this in star clusters and globular clusters. In part this is a natural result of fragmentation: in our model, these star clusters and globular clusters form {\it within} such baryonic building blocks. However, the picture cannot be quite this simple for two reasons. Firstly, we require that the phase space density of stars in our building blocks is low, yet in globular clusters and massive star clusters it is very high. Secondly, we have outlined the importance of supernovae winds in suppressing star formation on mass scales $\simlt 10^8$M$_\odot$. How then can star clusters with stellar mass $\simlt 10^5$M$_\odot$ ever form? 

\citet{2002MNRAS.336.1188K} provide some plausible answers to these problems. On the scale of star clusters, the star formation timescales are shorter than the lifetime of a typical O-star. This explains why gas expulsion from supernovae is not likely to be of great importance for star cluster formation (though it becomes very important for driving galactic scale winds and heating surrounding gas, as we have shown here). Instead, stellar winds and ionising flux from O-stars are the most important forms of internal feedback for star clusters forming with masses $\simlt 10^4$M$_\odot$. Above $10^4$M$_\odot$, supernovae become important for star clusters and can allow them to self enrich. Such massive clusters are still protected from destruction by supernovae winds, however, by their high density \citep{1989ApJ...339..171M}.

It is not clear how the very high gas densities required to form globular clusters can be achieved. Observational evidence suggests  a link to the star formation rate. In interacting systems (like galaxy mergers) and strong star bursts, the globular cluster and massive star cluster formation rate is much larger than in more quiescent systems (\bcite{1993AJ....106.1354W} and \bcite{1993MNRAS.264..611Z}). This view is also supported by recent theoretical work (see e.g. \bcite{2004ApJ...614L..29L}). It is clear that accurately modelling such effects is beyond the scope of this work. 

\section{Conclusions}\label{sec:conclusions}

Using analytic arguments and a suite of very high resolution ($\sim 10^3$\,M$_\odot$ per particle) cosmological hydro-dynamical simulations, we have argued that high redshift, $z \sim 10$, $M \sim 10^8$\,M$_\odot$ halos, form the smallest `baryonic building block' (BBB) for galaxy formation. These halos are just massive enough to efficiently form stars through atomic line cooling and to hold onto their gas in the presence of supernovae winds and reionisation. These combined effects, in particular that of the supernovae feedback, create a sharp transition: over the mass range $3-10 \times 10^7$M$_\odot$, the BBBs drop two orders of magnitude in stellar mass. Below $\sim 2 \times 10^7$M$_\odot$ galaxies will be dark with almost no stars and no gas. Above this scale is the smallest unit of galaxy formation: the BBB.

We have shown that supernovae feedback is important for these smallest galaxies. Not because it ejects the gas, but because it keeps the gas hot and extended. We find that the details of such feedback is not critical. Whether implemented as a heating term for gas surrounding star forming regions, or as a galaxy-wide wind, the results are similar. However, the combination of supernovae heating {\it and} supernovae driven galactic winds gives the best agreement with observations. Such feedback works by reducing the star formation efficiency in subhaloes. This keeps the surface density and rotational velocity of the stars which do form low. The smallest observed galaxies in the Local Group have very low surface brightness, as does the Milky Way old stellar halo. Without such supernovae feedback our model cannot reproduce these properties. Efficient cooling of the supernovae ejecta is prevented by the ionising background from reionisation.

We connected these BBBs to galaxies observed at $z=0$ using a mixture of linear theory arguments and results from other studies in the literature. In a galaxy the size of the Milky Way, O(100) such building blocks will be accreted. \citet{2005astro.ph.10370M} have recently shown that, of these, $\sim 10$\% survive to form the lowest mass Local Group dwarf galaxies. The remainder form the bulk of the Milky Way old stellar halo. The survivors will slowly form stars and lose gas over a Hubble time. Since neither reionisation nor supernovae winds actually eject gas from these galaxies, we require some other mechanism for this. Ram pressure stripping is a likely candidate. In this case, those BBBs on benign orbits which keep them far away from the Milky Way or Andromeda manage to retain their gas and slowly form stars - these become the lowest mass dwarf irregular galaxies (dIrr); those on more severe orbits lose their gas faster than they can form stars and become the dwarf spheroidals (dSphs). Both galaxy types become more metal rich and of lower abundance than the progenitor building blocks due to their more extended star formation. In this picture, the dSphs are `naked stellar halos', while the dIrrs have both an old, metal poor, stellar halo and a younger, more centrally concentrated population of stars. There is increasing observational evidence that this is indeed the case (\bcite{1996ApJ...467L..13M}, \bcite{2003Sci...301.1508M} and \bcite{2000AJ....119..177A}).

We have shown that the stars in our high redshift BBBs resemble in those in the faintest dSphs in the Local Group at the present epoch. This does not prove, however, that all of the Local Group dSphs formed in this way. \citet{2004ApJ...609..482K} and \citet{2005astro.ph..4277M} have suggested that some formed from quite different mechanisms and were much more massive in the past than at present. 

There should be many galaxies in the Local Group with surface brightness an order or two orders of magnitude fainter than those already found. We predict that these galaxies will have very similar total mass ($2-10 \times 10^7$\,M$_\odot$) to those satellites already discovered.

Finally, we have commented almost exclusively on the low-density Local Group, rather than the high-density cluster environment. This is mainly because the BBBs we form in our simulations are so small and faint that, at the moment, we can only hope to observe their $z=0$ counterparts in the very nearby universe. We expect, however, many such BBBs to survive to the present epoch, even in cluster environments. The continued accretion of such BBBs over a Hubble time should lead to an old metal-poor stellar halo being a ubiquitous feature of all large galaxies. However, it is important not to think of these BBBs as {\it the} building block for galaxy formation. We stress that they are the {\it smallest} building block and the major contributor to old stellar halos, not the bricks from which all galaxies are made. 

\section*{Acknowledgments.} 
The simulations were run on the COSMOS (SGI Altix 3700) supercomputer
at the Department of Applied Mathematics and Theoretical Physics in
Cambridge and on the Sun Linux cluster at the Institute of Astronomy in
Cambridge. COSMOS is a UK-CCC facility which is supported by HEFCE and
PPARC. AP would like to thank PPARC for the studentship. We would like to thank Andrea Maccio, Neil Trentham, Mark Wilkinson, Andrey Kravtsov, George Lake, Ben Moore, Lucio Mayer, Prasenjit Saha, Volker Springel and the anonymous referee for useful comments which led to this final version. We would like to thank Nick Gnedin for spotting an error in an earlier version. 

\bibliographystyle{mn2e}
\bibliography{/Users/justinread/Documents/LaTeX/BibTeX/refs}
 
\end{document}